\documentclass[journal]{IEEEtran}
\usepackage{cite}
\usepackage{hyperref}

\ifCLASSINFOpdf
   \usepackage[pdftex]{graphicx}
   \graphicspath{{../Figs/}}
\else
\fi
\usepackage{amsmath}
\usepackage{amssymb}
\usepackage{cleveref}
\crefrangeformat{equation}{(#3#1#4) -~(#5#2#6)}

\usepackage{amsthm}

\usepackage{array}

\ifCLASSOPTIONcompsoc
 \usepackage[caption=false,font=normalsize,labelfont=sf,textfont=sf]{subfig}
\else
 \usepackage[caption=false,font=footnotesize]{subfig}
\fi
\usepackage{url}

\usepackage[acronym]{glossaries}
\usepackage{color,soul}
\usepackage{breqn}
\usepackage{svg}
\usepackage{booktabs}
\usepackage{algorithm}
\usepackage{algpseudocode}

\newtheorem{theorem}{Theorem}

\theoremstyle{remark}
\newtheorem{remark}{Remark}

\ifCLASSOPTIONcompsoc
    \usepackage[caption=false, font=normalsize, labelfont=sf, textfont=sf]{subfig}
\else
\usepackage[caption=false, font=footnotesize]{subfig}
\fi
\usepackage{tabularx}
\newcolumntype{C}[1]{>{\centering\let\newline\\\arraybackslash\hspace{0pt}}m{#1cm}}
\usepackage{siunitx}
\usepackage{multirow} 

\begin{document}
%
\title{Exploiting Convexity of Neural Networks in Dynamic Operating Envelope Optimization for Distributed Energy Resources}
%
%
%

\author{Hongyi Li,~\IEEEmembership{Member,~IEEE,}
Liming Liu,~\IEEEmembership{Graduate Student Member,~IEEE,}
Yunyi Li,
Zhaoyu Wang,~\IEEEmembership{Fellow,~IEEE,}
\thanks{	

H. Li, L. Liu, Y. Li, and Z. Wang are with the Department of Electrical and Computer Engineering, Iowa State University, Ames, IA 50011 USA (email: hongyili@iastate.edu; limingl@iastate.edu; liyunyi@iastate.edu; wzy@iastate.edu). (Corresponding author: \textit{Zhaoyu Wang}.)
	}
}

\maketitle

\begin{abstract}
The increasing penetration of distributed energy resources (DERs) brings opportunities and challenges to the operation of distribution systems. To ensure network integrity, dynamic operating envelopes (DOEs) are issued by utilities to DERs as their time-varying export/import power limits. Due to the non-convex nature of power flow equations, the optimization of DOEs faces a dilemma of solution accuracy and computation efficiency. To bridge this gap, in this paper, we facilitate DOE optimization by exploiting the convexity of input convex neural networks (ICNNs). A DOE optimization model is first presented, comprehensively considering multiple operational constraints. We propose a constraint embedding method that allows us to replace the non-convex power flow constraints with trained ICNN models and convexify the problem. To further speed up DOE optimization, we propose a linear relaxation of the ICNN-based DOE optimization problem, for which the tightness is theoretically proven. The effectiveness of the proposed method is validated with numerical case studies. Results show that the proposed ICNN-based method outperforms other benchmark methods in optimizing DOEs in terms of both solution quality and solution time.
\end{abstract}


%
\IEEEpeerreviewmaketitle

\section{Introduction}
\IEEEPARstart{T}{he} penetration level of distributed energy resources (DERs), such as photovoltaics (PV) panels, battery energy storage systems, and responsive loads, is continuously growing in distribution systems~\cite{wang_networked_2016}. As cost-efficient and low-carbon energy resources, DERs can significantly reduce the operation costs and carbon emissions of the distribution system~\cite{li_decentralized_2023}. According to FERC Order 2222, DERs are encouraged to actively participate in electricity markets alongside traditional generation resources, which further broadens the benefits brought by DERs \cite{federalenergyregulatorycommission_participation_2020}. However, besides opportunities, DERs also pose significant challenges to the operation of distribution systems~\cite{wang_decentralized_2016}.

Although the impact of a single DER is limited, aggregated DERs can have a nontrivial influence on the voltage profile and power flows of the distribution system \cite{he_leasing_2025}. Uncoordinated DER aggregations could lead to multiple operational issues, including over-/under-voltages, overloading of lines and transformers, and reverse power flow, which will consequentially result in accelerated insulation degradation and protection device malfunctions~\cite{jang_Probabilistic_2024}. Thus, it is necessary for system operators and utilities to coordinate the behaviors of aggregated DERs to ensure network integrity \cite{wang_coordinated_2015}. \textcolor{black}{Both centralized~\cite{meena_optimisation_2019} and distributed~\cite{li_consensus-based_2023-2} DER coordination mechanisms are proposed by researchers. However, in the context of third-party-owned DERs, coordination is intricate since utilities cannot directly intervene with DER behavior or share distribution feeder models~\cite{mahmoodi_capacity_2024}.}

To ensure network integrity without directly controlling DER power, the concept of dynamic operating envelopes (DOEs) is proposed \cite{moring_Nodal_2024}. DOEs are time-varying export/import power limits issued by utilities to individual DERs, based on the available capacity of the distribution systems \cite{liu_grid_2021}. DOEs are widely adopted by utilities to allow DERs to participate in local energy markets \cite{hoque_dynamic_2023} and demand response programs \cite{lankeshwara_development_2025}, without threatening the operational constraints of distribution systems. \textcolor{black}{However, the central challenge of DOE optimization is the non-convex and unbalanced characteristics of the distribution system power flow. One possible approach to addressing this issue is to calculate DOEs using an iterative simulation method, which is similar to evaluating hosting capacity~\cite{azim_dynamic_2024}. For example, in Ref.~\cite{liu_computation_2024}, the initial DOEs are set to large values. Then, through iterative power flow simulations, the upper and lower boundaries of DOEs are heuristically adjusted until all constraint violations are eliminated. While this iterative process is easy to integrate into existing grid management systems, ensuring solution optimality is challenging due to its discrete and heuristic nature. Besides optimality concern, DOEs are usually updated frequently and with fine temporal granularity, which requires efficient solution algorithms~\cite{bassi_final_2023}. The increasing DER penetration and system scale also pose great challenges to the scalability of DOE optimization algorithms~\cite{avramidis_comprehensive_2021}. Therefore, the iterative simulation method often falls short when numerous DERs are considered in a large-scale system.}

\textcolor{black}{To handle increasing DER penetration in distribution systems, researchers investigate optimization-based DOE calculation methods. In these methods, convex relaxations are adopted to make power flow constraints tractable. Second-order cone relaxation is adopted in Ref.~\cite{moring_Inexactness_2023} for DOE optimization, while the authors also point out the inexactness of relaxation when the objective function is not monotonic. In Ref. \cite{nazir_gridaware_2022}, researchers derive a second-order inner approximation of the power flow relationship to support DOE optimization. However, the coefficients are derived based on nominal operating points. Another broadly adopted convexification technique is to linearize the power flow constraints. In Ref.~\cite{alam_allocation_2024a}, the linear power-voltage sensitivity is derived and used to optimize DOE with voltage constraints. Similarly, in Ref. \cite{liu_using_2022,liu_linear_2024a,gao_equitable_2025,jiang_bargainingbased_2025}, researchers optimize DOEs with linearized power flow models. Linearized power flow models are efficient for optimization and provide reasonable accuracy when operating conditions remain relatively stable. However, DOE optimization always involves a wide range of loading and penetration levels, making accuracy a main concern. In summary, most existing model-based DOE optimization methods entail an inherent trade-off between accuracy and computational efficiency.}

\textcolor{black}{With advancements in artificial intelligence and big data technologies, data-driven optimization methods are gaining increasing attention and are seen as promising alternatives to model-based methods. In Ref. \cite{liu_DataDriven_2019}, regression models are proposed to approximate the relationship between bus power, voltage, and power flows, based on smart meter data. Similar ideas are applied in Ref. \cite{kumarawadu_smart_2025} to generate a linear power flow surrogate for DOE optimization. Although data-driven regression methods eliminate the assumption of nominal operating points, neglecting non-linearity may cause issues. In the broader literature, neural networks have been utilized to support optimal power flow.} In Ref. \cite{chen_scheduling_2021}, the authors point out that multilayer perceptrons (MLPs) can learn and replace intractable power flow constraints with mixed-integer linear counterparts. Although mixed-integer linear programming (MILP) problems can be solved to optimality, as the distribution system scales up, the number of integer variables also increases, resulting in a significant increase in computational burden \cite{chen_efficient_2024}. Different from MLPs, input convex neural networks (ICNNs) can be reformulated into a group of non-linear convex functions \cite{amos_input_2017}, making it appealing for downstream optimization tasks, such as optimal power flow \cite{zhang_convex_2022,cheng_input_2024}. In Ref. \cite{wu_transient_2024}, the authors use an ICNN to model the transient stability index, which is relaxed into a group of linear constraints, enhancing optimization efficiency. \textcolor{black}{However, strategies for handling constrained ICNN outputs with both upper and lower limits are lacking. Additionally, the method adopted in Ref. \cite{wu_transient_2024} cannot adapt to changes in the constraint parameter, which could result in expensive retraining when the utility wants to modify the operational limits. In summary, research efforts need to be dedicated to the development of accurate, adaptive, and computationally efficient DOE optimization methods.}

In this paper, we study the DOE optimization problem and address the research gaps by exploiting the convexity of ICNNs. The optimization model for DOE is first established, incorporating various operational constraints that align with utility needs. \textcolor{black}{Then, we leverage ICNNs to achieve time-efficient and high-accuracy DOE optimization. By reformulating the optimization problem, we propose a constraint embedding method that utilizes ICNNs to convexify the distribution power flow model, considering constrained outputs with both upper and lower bounds. The ICNN-aided DOE optimization model is further relaxed into a linear programming problem, significantly improving the computational efficiency. The equivalence between the original problem and the reformulation is theoretically proved and demonstrated by experiments. To further reduce the computational burdens of ICNN training and DOE optimization, we propose a constraint retrenchment strategy based on the structure of the distribution network.}
Case studies are conducted on the IEEE 123-node test feeder and the EPRI Ckt5 test feeder to validate the effectiveness and scalability of the proposed methods. Results show that the proposed ICNN-based method achieves satisfactory improvements on both accuracy and computation efficiency compared with benchmark methods. The main contributions of this paper are:
\begin{enumerate}
    \item We develop a comprehensive DOE optimization model for DERs in the distribution system, considering voltage limits, thermal capacities, and reverse power flow limits. After reformulating the DOE optimization problem, the operational constraints are handled by individual ICNN models in a modularized way.
    \item We propose a constraint embedding method to encode operational constraints as a violation-regularization layer of the trained ICNN, \textcolor{black}{which effectively handles both upper and lower bounds of operational constraints.} In this way, ICNNs can learn the physical quantities of distribution systems, \textcolor{black}{thereby improving the adaptivity of constraint parameter changes.}
    \item With the constraint embedding method, the ICNNs are translated into linear inequality constraints, replacing the intractable non-convex power flow in the DOE optimization problem. The proposed ICNN-based DOE optimization method significantly reduces computation complexity without compromising accuracy.
    \item \textcolor{black}{To further reduce computation burdens in ICNN training and lower the dimensionality of the reformulated DOE optimization problem, we propose a constraint retrenchment strategy to prune out unbinding constraints and simplify ICNN output dimensions based on the distribution system structure.}
\end{enumerate}

The remainders of this paper are organized as follows: In Section~\ref{section 2}, the DOE optimization problem is formulated. The proposed ICNN-based method is presented in Section~\ref{section 3}. In Section~\ref{section 4}, the effectiveness of the proposed DOE optimization method is demonstrated through numerical case studies. Finally, Section~\ref{section 5} concludes the paper.

\section{Dynamic Operating Envelopes for Distributed Energy Resources}
\label{section 2}
In this section, we first introduce the framework for calculating DOEs in distribution systems. Then, we present the DOE optimization problem for DERs and reformulate the problem for ICNN-based solving.

\subsection{Dynamic Operating Envelope Calculation Framework}
\begin{figure}
    \centering
    \includegraphics[width=\linewidth]{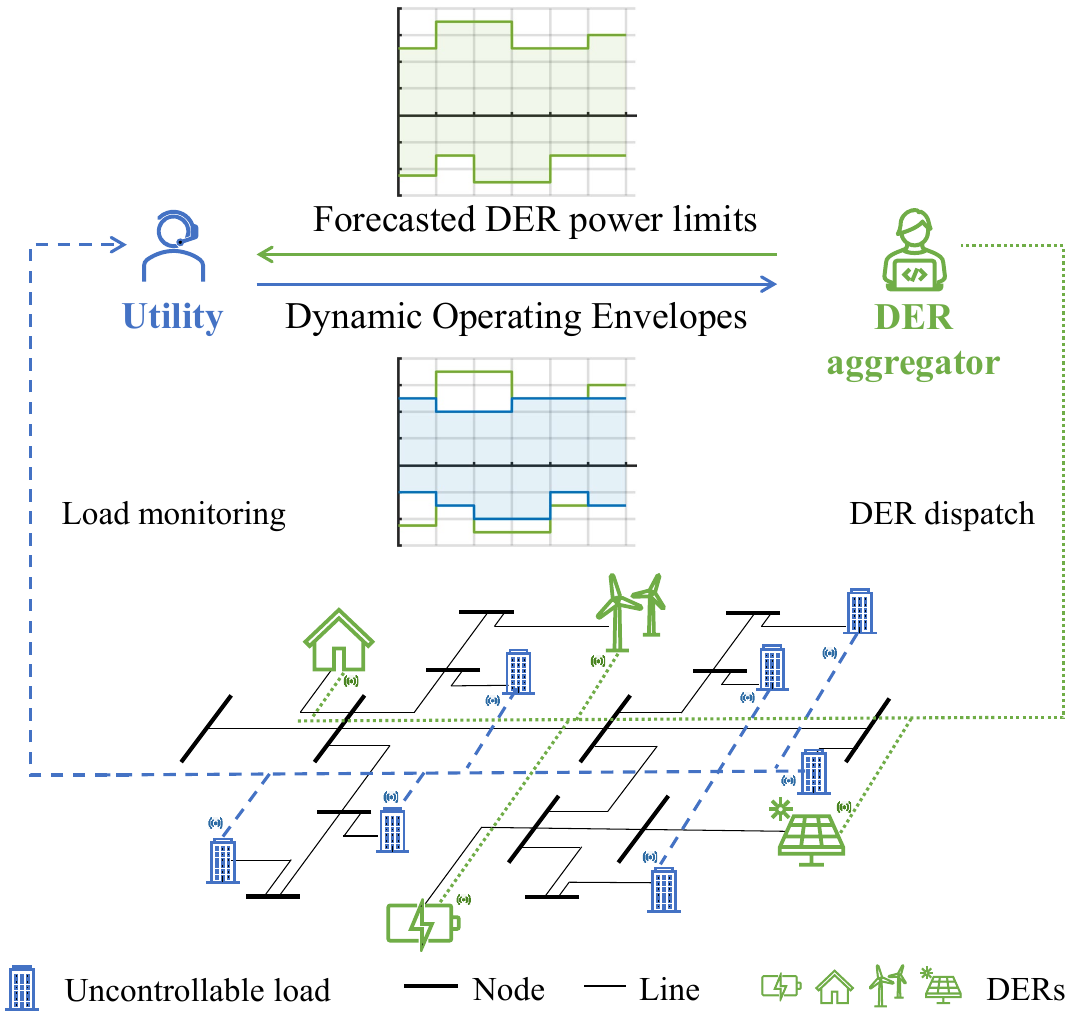}
    \caption{DOE optimization framework for DERs in the distribution system.}
    \label{DOE framework}
\end{figure}
In the distribution system, there are uncontrollable loads and DERs. We assume that with load monitoring devices, utilities can accurately estimate the power consumption trends of uncontrollable loads. DERs are managed by DER aggregators, which are responsible for forecasting the available power of DERs. The calculation of DOEs for DERs involves two-way communications. As shown in Fig.~\ref{DOE framework}, with the forecasts, DER aggregators send prospective upper and lower DER power limits to the utility\footnote{\textcolor{black}{The main objective of this paper is to develop accurate and efficient DOE optimization methods for utilities. Although the limits provided by DERs would affect the solution, the effectiveness of the proposed methods does not depend on the limits. The integrity of these limits can be guaranteed by specifying requirements in the interconnection agreement. In this paper, we assume that DERs can predict their power limits and are willing to share these actual limits with the utility.}}. \textcolor{black}{Here, a negative lower limit represents the maximum import power of a DER. The utility examines the upper and lower limits based on anticipated load levels and determines DOEs for each DER. Here, the utility only reduces the upper limit or increases the lower limit to ensure that no operational constraint is breached.} Then, DOEs are issued to DER aggregators and used to constrain the power of individual DERs.

\subsection{DOE Optimization Problem}
\label{section 2b}
Before presenting the DOE optimization problem, we outline some necessary notations first. In this paper, we consider a radial distribution network and use $i,j$ as the node indexes. The set of nodes is $\mathcal{M}$. The direct downstream nodes of node $j$ are included in set $\mathcal{N}_j^-$. The distribution systems are typically three-phase unbalanced. We use $\phi,\psi\in\Phi=\{a,b,c\}$ to index the phases of the nodes. The line set is denoted by $\mathcal{E}$, where $(i,j)\in \mathcal{E}$ means there is a line connecting nodes $i$ and $j$. For each line, we use $r_{ij,\phi\psi}$ and $x_{ij,\phi\psi}$ to denote the self/mutual resistance and reactance between different phases of node $i$ and node $j$. $P_{ij,\phi}$, $Q_{ij,\phi}$ and $\left| I_{ij,\phi} \right|$ are the active and reactive power flow, and current magnitude flowing through phase $\phi$ of line $(i,j)$. With a slight abuse of symbols, we also use $j$ to index the DER connected to node $j$. Symbols $t$ and $\mathcal{T}$ denote the index and the set of time intervals in the DOE optimization problem. As discussed in the previous subsection, for each time interval $t$, we assume that the DER aggregator forecasts the upper and lower active power limits of DER $j$ at phase $\phi$, denoted by $p_{j,\phi,t}^{\max}$ and $p_{j,\phi,t}^{\min}$. For buses without DER, we set $p_{j,\phi,t}^{\max}=p_{j,\phi,t}^{\min}=0$. The reactive power of each DER $q_{j,\phi,t}^{\text{DER}}$ is assumed to be constant. Both active and reactive power of DERs are modeled as power injections. The active and reactive power of uncontrollable loads monitored by the utility are $p_{j,\phi,t}^{0}$ and $q_{j,\phi,t}^{0}$, respectively. To make the discussions in Section~\ref{section 3} clear and easy to understand, we define vectors $\boldsymbol{V}_t$, $\boldsymbol{I}_t$ and $\boldsymbol{P}_t$. Vector $\boldsymbol{V}_t$ concatenates the voltage magnitudes $|V_{j,\phi,t}|$ of all nodes at time $t$, while vectors $\boldsymbol{I}_t$ and $\boldsymbol{P}_t$ consist of current magnitudes $|I_{ij,\phi,t}|$ and active powers $P_{ij,\phi,t}$ of each line $(i,j)$ in the distribution system.

Based on DERs' power limit forecasts and the load level, the utility solves an optimization problem to determine DOEs for DERs. \textcolor{black}{Taking the upper limit $p_{j,\phi,t}^{+}$ as an example}, the DOE optimization problem is formulated as follows:
\begin{subequations}
\label{DOE optimization problem}
\begin{align}
\label{DOE obj}
	\color{black}{\underset{p_{j,\phi,t}^{+}}{\text{min}}}\;\;\;\;&\color{black}{w^{\text{DOE}}\sum_{t\in \mathcal{T}}{\sum_{j\in \mathcal{M} ,\phi \in \Phi}{\left| p_{j,\phi ,t}^{\max}-p_{j,\phi ,t}^{+} \right|}}+\sum_{t\in \mathcal{T}}{w^{\text{loss}}P_{t}^{\text{loss}}}}\\
    \label{DER power constraint}
	\color{black}{\text{s}.\text{t}.}\;\;\;\;&\color{black}{p_{j,\phi,t}^{\min}\le p_{j,\phi,t}^{+}\le p_{j,\phi,t}^{\max},}\\
    \label{bus power definition}
	&\color{black}{p_{j,\phi,t}=p_{j,\phi,t}^{0}-p_{j,\phi,t}^{+}, q_{j,\phi,t}=q_{j,\phi,t}^{0}-q_{j,\phi,t}^{\text{DER}},}\\
	&\color{black}{P_{ij,\phi ,t}=\sum_{k\in \mathcal{N} _{j}^{-}}{P_{jk,\phi ,t}}+p_{j,\phi ,t}\notag}\\
    \label{active power flow}
    &\color{black}{\quad\quad\quad\quad\quad\quad\quad\quad+\sum_{\psi \in \Phi}{r_{ij,\phi \psi}\left| I_{ij,\psi ,t} \right|^2},}\\
	&\color{black}{Q_{ij,\phi ,t}=\sum_{k\in \mathcal{N} _{j}^{-}}{Q_{jk,\phi ,t}}+q_{j,\phi ,t}\notag}\\
    \label{reactive power flow}
    &\color{black}{\quad\quad\quad\quad\quad\quad\quad\quad+\sum_{\psi \in \Phi}{x_{ij,\phi \psi}\left| I_{ij,\psi ,t} \right|^2},}\\
	&\color{black}{\left| V_{j,\phi ,t} \right|^2=\left| V_{i,\phi ,t} \right|^2+\sum_{\psi \in \Phi}{\left( r_{ij,\phi \psi}^{2}+x_{ij,\phi \psi}^{2} \right) \left| I_{ij,\psi ,t} \right|^2} \notag}\\
    \label{voltage drop}
    &\color{black}{\quad\quad-2\sum_{\psi \in \Phi}{\left( r_{ij,\phi \psi}P_{ij,\psi ,t}-2x_{ij,\phi \psi}Q_{ij,\psi ,t} \right)},}\\
    \label{current definition}
	&\color{black}{\left| I_{ij,\phi ,t} \right|^2=\frac{P_{ij,\phi ,t}^{2}+Q_{ij,\phi ,t}^{2}}{\left| V_{i,\phi ,t} \right|^2},}\\
    \label{voltage limit}
    &\boldsymbol{V}^\min \le \boldsymbol{V}_t \le \boldsymbol{V}^\max,\\
    \label{thermal limit}
    &\boldsymbol{I}_t\le \boldsymbol{I}^\max,\\
    \label{reverse power flow limit}
    &\boldsymbol{P}_t\ge \boldsymbol{P}^\min.
\end{align}
\end{subequations}
The objective function (\ref{DOE obj}) includes two terms. The first term minimizes the deviation of DOE from the original upper limits, and the second term aims to reduce the distribution system's active power loss, denoted by $P^{\text{loss}}_{t}$. \textcolor{black}{In this paper, we include the minimization of the active power loss as a secondary objective, which is usually considered an important index for distribution system operation efficiency. In practice, the utility can set weighting factors $w_{\text{DOE}}=1$ and $w_{\text{loss}}=0$ to purely focus on optimizing DOEs. Other combinations of weighting factors can be explored according to the emphasize of the distribution system operation.} Constraint (\ref{DER power constraint}) ensures that DOE lies within the original upper and lower limits. Constraint (\ref{bus power definition}) defines the net active and reactive loads of phase $\phi$ of node $j$. Constraints (\ref{active power flow}) - (\ref{current definition}) are the DistFlow model describing the power flow equations in a radial distribution system~\cite{baran_optimal_1989,gan_convex_2014}. Note that Constraint (\ref{current definition}) introduces non-convexity to the DOE optimization problem. Constraints (\ref{voltage limit}) - (\ref{reverse power flow limit}) are element-wise inequalities. Constraint (\ref{voltage limit}) avoids under- and over-voltage issues by limiting bus voltages within lower and upper bounds $\boldsymbol{V}^\min$ and $\boldsymbol{V}^\max$. Constraint (\ref{thermal limit}) guarantees that the thermal limits $\boldsymbol{I}^\max$ of the lines are not breached. Defining $P_{ij,\phi}>0$ as the active power flowing from upstream node $i$ to downstream node $j$ on phase $\phi$, constraint (\ref{reverse power flow limit}) limits the reverse power flow. The reverse power flow limits $\boldsymbol{P}^\min$ are determined by utilities according to the configurations of legacy devices, such as overcurrent relays. 

\begin{remark}
    \textcolor{black}{In this paper, our main focus is to develop tractable counterparts for non-convex power flow equations (\ref{active power flow}) - (\ref{current definition}). Note that there is no temporal coupling in equations (\ref{active power flow}) - (\ref{current definition}). To make the discussions simple and easy to understand, we present our proposed method by optimizing the subproblems of each time interval $t$ independently. Thus, in the following discussion, we omit the time index $t$ for simplicity. However, in practice, temporal coupling could be introduced to the DOE optimization problem, for example, by utility-scale battery energy storage systems. In those scenarios, our proposed methods are directly applicable, as they only deal with power flow equations (\ref{active power flow}) - (\ref{current definition}).}
\end{remark}

In optimization problem (\ref{DOE optimization problem}), we formulate operational constraints (\ref{voltage limit}) - (\ref{reverse power flow limit}) as hard constraints. However, in practice, utilities usually allow temporary violation of these constraints, but want to avoid persistent violations. Thus, we reformulate problem (\ref{DOE optimization problem}) by relaxing operational constraints and adding them as penalty terms in the objective function:
\begin{subequations}
\label{reformulated DOE optimization}
    \begin{align}
    \label{reformulated DOE obj}
	\underset{p_{j,\phi}^{+}}{\text{min}}\;\;\;\;&w^{\text{DOE}}\sum_{j\in \mathcal{M} ,\phi \in \Phi}{\left| p_{j,\phi ,t}^{\max}-p_{j,\phi ,t}^{+} \right|}+\sum_{t\in \mathcal{T}}{w^{\text{loss}}P_{t}^{\text{loss}}} \notag \\
    &\quad\quad\quad\quad\quad\quad\quad\quad+w^{\text{v}}\delta ^{\text{v}}+w^{\text{ol}}\delta ^{\text{ol}}+w^{\text{rpf}}\delta^{\text{rpf}}\\
	\text{s}.\text{t}.\;\;\;\;& (\ref{DER power constraint}) - (\ref{current definition}),
\end{align}
\end{subequations}
where $\delta^{\text{v}}$, $\delta^{\text{ol}}$ and $\delta^{\text{rpf}}$ are penalty terms for violating constraints (\ref{voltage limit}), (\ref{thermal limit}) and (\ref{reverse power flow limit}). Weighting factors $w^{\text{v}}$, $w^{\text{ol}}$, and $w^{\text{rpf}}$ are used to emphasize different constraints and scale the quantities to comparable levels. The constraint violations are defined as:
\begin{align}
\label{delta v definition}
    &\delta^{\text{v}}=\mathbf{1}^{\text{T}}\left[\max \left( \boldsymbol{V}-\boldsymbol{V}^{\max},0 \right) +\max \left( \boldsymbol{V}^{\min}-\boldsymbol{V},0 \right)\right] ,\\
\label{delta ol definition}
    &\delta^{\text{ol}}=\mathbf{1}^{\text{T}}\max \left( \boldsymbol{I}-\boldsymbol{I}^{\max},0 \right) ,\\
\label{delta rpf definition}
    &\delta^{\text{rpf}}=\mathbf{1}^{\text{T}}\max \left( \boldsymbol{P}^{\min}-\boldsymbol{P},0 \right),
\end{align}
\textcolor{black}{where the operator $\max(\cdot)$ conducts element-wise comparison. Left-multiplying $\mathbf{1}^{\text{T}}$ sums up the output of the $\max(\cdot)$ operator. Thus, $\delta^{\text{v}}$. $\delta^{\text{ol}}$ and $\delta^{\text{rpf}}$ correpsond to the sum of constraint violations (\ref{voltage limit}), (\ref{thermal limit}) and (\ref{reverse power flow limit}) across the system, respectively.}

\begin{remark}
    \textcolor{black}{The penalized optimization problem (\ref{reformulated DOE optimization}) is a relaxation of the original problem (\ref{DOE optimization problem}). The relaxation ensures that the optimization problem is always feasible. The penalty factors $w^{\text{v}}$, $w^{\text{ol}}$ and $w^{\text{rpf}}$ must be carefully selected to ensure the equivalence of the penalized problem and the original constrained one. As discussed in Ref.~\cite{cherukuri_distributed_2015, boyd_convex_2004}, the solutions of the penalized problem and the original problem coincide when the penalty factors are sufficiently large. In practice, for some distribution systems, temporary and minor violations of operational constraints (such as slightly overloading lines) are considered acceptable if they occur infrequently~\cite {he_optimization_2021}. Thus, if the utility wants to strictly avoid constraint violation, it can use large penalty factors, such that the penalized form (\ref{reformulated DOE optimization}) yields the same solution. Otherwise, our proposed formulation allows the utility an additional degree of freedom to adapt the optimization according to its risk appetite.}
\end{remark}

\begin{remark}
    \textcolor{black}{The optimization of lower bounds $p_{j,\phi}^-$ of DOEs can be achieved by replacing the first term of objective (\ref{reformulated DOE obj}) with $w_{\text{DOE}}\sum_{j\in \mathcal{M},\phi\in\Phi}{\left| p_{j,\phi}^{-}-p_{j,\phi}^{\min} \right|}$. The DOE optimization methods proposed in Section~\ref{section 3} are agnostic to the first term of objective (\ref{reformulated DOE obj}) and applicable to both $p_{j,\phi}^{+}$ and $p_{j,\phi}^-$ optimizations.}
\end{remark}

\section{Input Convex Neural Network for DOE Optimization Problem}
\label{section 3}
In this section, we first introduce some preliminaries of ICNN. Then, we propose the constraint embedding method for ICNN and discuss how to translate the trained ICNN models into linear inequality constraints. To further reduce computation burdens, we propose a constraint retrenchment method based on the structure of distribution systems.

\subsection{Preliminaries of Input Convex Neural Networks}
\subsubsection{Mathematic Model of ICNN}
\begin{figure}
    \centering
    \includegraphics[width=\linewidth]{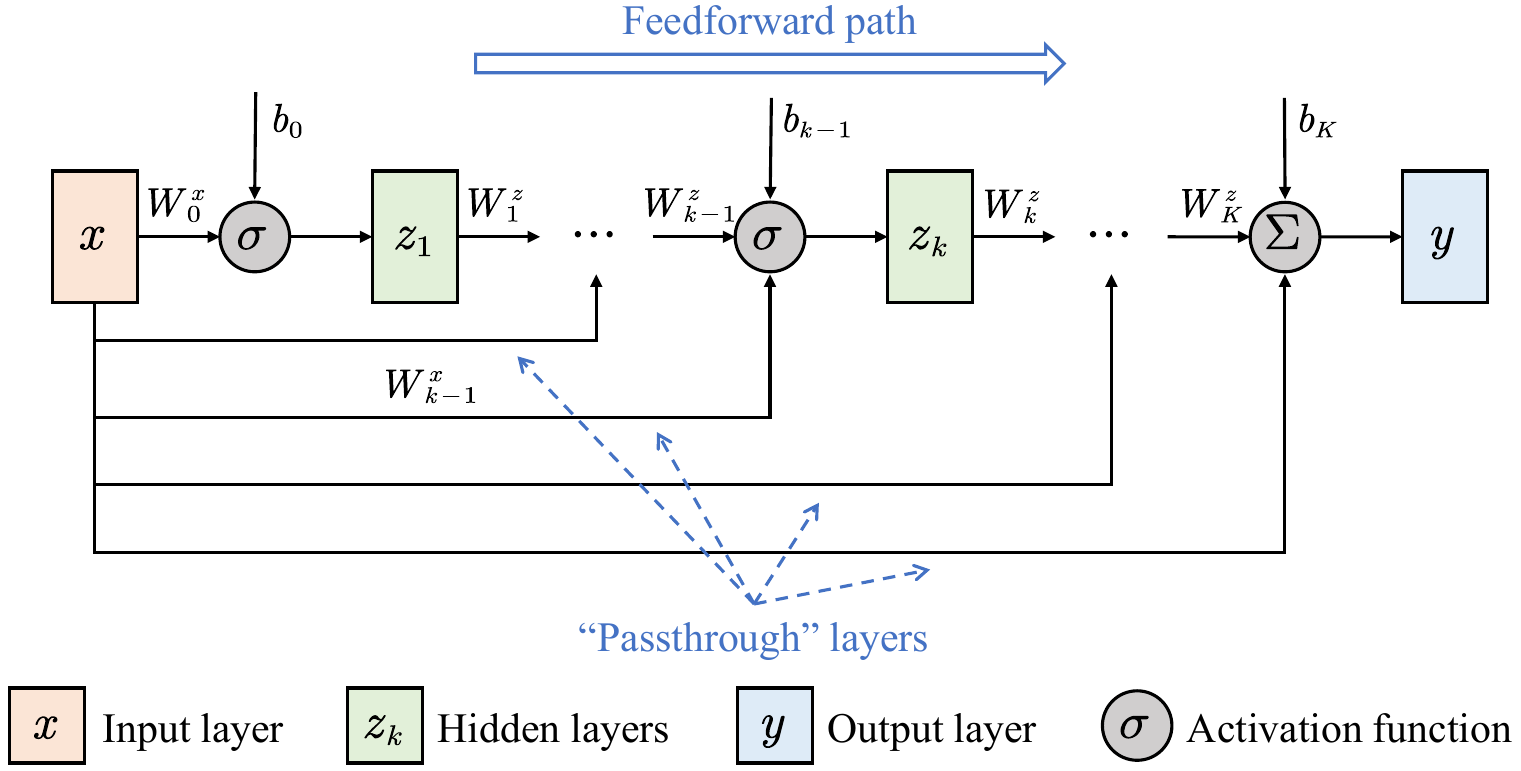}
    \caption{Illustration of ICNN.}
    \label{ICNN}
\end{figure}
Proposed in Ref.~\cite{amos_input_2017}, ICNNs are neural networks that can be interpreted as convex functions of their inputs. As shown in Fig.~\ref{ICNN}, in addition to the feedforward path, an ICNN has ``passthrough" layers from input $x$ to hidden layers, which distinguishes it from other neural networks. For an ICNN with $K$ hidden layers, the mathematical formulation is:
\begin{subequations}
    \begin{align}
    &z_1=\sigma\left(W_{0}^{x}x+b_{0}\right),\\
	&z_{k+1}= \sigma\left( W_{k}^{z}z_{k}+W_{k}^{x}x+b_{k}\right), k=1,\cdots,K-1,\\
    &y=W_{K}^{z}z_{K}^{l}+W_{K}^{x}x+b_{K},
\end{align}
\end{subequations}
where $x$ and $y$ are the input and output vectors of ICNN; $z_{k}$ is the output of hidden layers; $W_{k}^{z}$, $W_{k}^{x}$ and $b_{k}$ are the weights and biases of ICNN; $\sigma(\cdot)$ are the activation functions. Using $\theta$ to represent all the weights and biases, the output $y$ can be written as $y=f(x;\theta)$, which is a convex function of $x$ given that the activation functions $\sigma(\cdot)$ are convex and $z_{k}$ have non-negative weights $W_{k}^{z}\ge0,k=1,\cdots,K-1$. Although non-negativity constraints are added to $W_{k}^{z}$, the value of $W_{k}^{x}$ and $b_{k}$ are not restricted, allowing ICNN to have substantial representation power for complex input-output relationships.

\subsubsection{Exact Inference of ICNN}
Inference is the process of obtaining the output of the trained neural network by feeding a new input. Conventionally, neural network inference is completed through algebraic operations. For ICNN, when using the rectified linear unit (ReLU) as the activation function, its exact inference can be equivalently written as a linear programming (LP) problem. Since solving optimization problems is typically more time-consuming than algebraic operations, the optimization-based inference of ICNN is rarely used to obtain outputs. However, this feature is appealing when ICNNs are integrated into optimization problems. To exploit the convexity and linearity of ICNN, we adopt ReLU as the activation function throughout the remainder of this paper. The ReLU activation function is defined as:
\begin{align}
\label{ReLU}
    \text{ReLU}(x)=\max(x,0).
\end{align}
An ICNN with $K$ hidden layers and ReLU activation functions can be written as:
\begin{align}
\label{z1o}
	&z_{1}=\max \left( W_{0}^{x}x+b_{0},0 \right) ,\\
\label{zko}
	&z_{k+1}=\max \left( W_{k}^{z}z_{k}+W_{k}^{x}x+b_{k},0 \right), k=1,\cdots,K-1,\\
\label{yo}
	&y=W_{K}^{z}z_{K}+W_{K}^{x}x+b_{K}.
\end{align}
Applying the above inference process is equivalent to solving the following LP problem:
\begin{subequations}
\label{exact inference of ICNN}
\begin{align}
\underset{z_k,y}{\text{min}}\;\;\;\;&y\\
\text{s.t.}\;\;\;\; &z_1\ge W_{0}^{x}x+b_0,\\
&z_{k+1}\ge W_{k}^{z}z_k+W_{k}^{x}x+b_k, k=1,\cdots ,K-1,\\
&z_k\ge 0, k=1,\cdots ,K,\\
&y\ge W_{K}^{z}z_{K}^{l}+W_{K}^{x}x+b_K.
\end{align}
\end{subequations}
In the following subsections, we discuss how to leverage the convexity and linearity of ReLU-activated ICNNs, which can be effectively solved by both commercial solvers and open-source solvers.

\subsection{Learning Power Flow Constraints with ICNN}
The key challenge in solving problem (\ref{reformulated DOE optimization}) is to handle power flow constraints (\ref{active power flow}) - (\ref{current definition}) effectively. We assume that utilities have historical operation data, including bus power injections, voltages, line flows, and losses. To facilitate the solving of problem (\ref{reformulated DOE optimization}), we use ICNNs to learn the mapping from bus power injections to the power loss term and the constraint violation terms. With the proposed methods, we replace the power flow constraints with convex and linear counterparts.

In existing literature, researchers usually let neural networks learn the safe distance to constraint boundaries to guarantee the convexity of the replicated constraints\cite{chen_scheduling_2021,cheng_input_2024}. However, since the calculation of safe distances already involves specific constraint parameters, such as voltage limits, the trained neural networks cannot adapt to parameter changes. To improve the adaptability of the proposed method, we train ICNNs to directly learn the mapping from bus power to active power loss, node voltages, line current magnitude, and line active power, respectively. 

To simplify notations, we define an superscript $o\in \{\text{loss,v,ol,rpf}\}$ to distinguish and index the ICNNs (\ref{z1o}) - (\ref{yo}) that handle the power loss term, the voltage constraint (\ref{delta v definition}), the overloading constraint (\ref{delta ol definition}) and the reverse power flow constraint (\ref{delta rpf definition}). For the four ICNNs, we define the same input  $x=[p_{j,\phi};q_{j,\phi}]$, which is a column vector concatenating the net active and reactive loads of all buses. The definition of the output vector $y^o$ is heterogeneous, which is discussed in detail in the following paragraphs. For simplicity of formulation, we do not specify the number of hidden layers and the number of neurons in each layer of the ICNN, as our proposed methods are agnostic to these configurations.

Since the active power loss $P^{\text{loss}}$ is not constrained by upper or lower limits, it is straightforward to set $y^{\text{loss}}=P^{\text{loss}}$. However, for the penalty terms $\delta^{\text{v}}$, $\delta^{\text{ol}}$ and $\delta^{\text{rpf}}$, the corresponding ICNN outputs need to be carefully designed. Otherwise, we cannot exploit the convexity of ICNNs in the optimization problem. Here, we take the voltage limit constraint (\ref{voltage limit}) as an example and discuss the design of $y^{\text{v}}$. Assume that we train an ICNN that has an output $y^{\text{v}}=\boldsymbol{V}=f(x|\theta^{\text{v}})$. According to the discussion in the last subsection, $y^{\text{v}}$ is a convex function of input $x$. Substituting $\boldsymbol{V}=f(x|\theta^{\text{v}})$ into equation (\ref{delta v definition}) yields:
\begin{align}
    &\delta _{\text{v}}=\mathbf{1}^{\text{T}}\left[ \max \left( f(x|\theta^{\text{v}})-\boldsymbol{V}^{\max},0 \right) \right. \notag\\
    &\quad\quad\quad\quad\quad\quad\quad\quad\left.+\max \left( \boldsymbol{V}^{\min}-f(x|\theta^{\text{v}}),0 \right) \right].
\end{align}

It is obvious that the first term is convex on $x$, as it involves the positive linear combinations of $f(x|\theta^{\text{v}})$ and a max operator. However, due to the negative sign of $f(x|\theta^{\text{v}})$, the second term becomes a non-convex function on $x$. Thus, if the ICNN only learns the physical quantities $\boldsymbol{V}$, $\boldsymbol{I}$ and $\boldsymbol{P}$, we can only handle upper limits in constraints (\ref{voltage limit}) - (\ref{reverse power flow limit}). For the lower limits, we cannot reformulate the ICNNs into convex counterparts of the power flow equations. To address this issue, we treat upper and lower limits heterogeneously and use separate ICNN outputs to calculate penalty terms. We define the following ICNN outputs:
\begin{align}
\label{y definition}
    y^{\text{v}}=\left[ \begin{array}{c}
	\boldsymbol{V}\\
	-\boldsymbol{V}\\
\end{array} \right] , y^{\text{ol}}=\boldsymbol{I}, y^{\text{rpf}}=-\boldsymbol{P},
\end{align}
and the corresponding vectors of constraint parameters:
\begin{align}
\label{lim definition}
    \varepsilon^{\text{v}}=\left[ \begin{array}{c}
	-\boldsymbol{V}^{\max}\\
	\boldsymbol{V}^{\min}\\
\end{array} \right] , \varepsilon^{\text{ol}}=-\boldsymbol{I}^{\max} , \varepsilon^{\text{rpf}}=\boldsymbol{P}^{\min}.
\end{align}

We take the voltage constraint (\ref{voltage limit}) as an example again to explain the rationales behind definitions (\ref{y definition}) and (\ref{lim definition}). Substituting $y^{\text{v}}=f(x|\theta^{\text{v}})$ into equation (\ref{delta v definition}) yields:
\begin{align}
    \delta^{\text{v}}=&\mathbf{1}^{\text{T}}\max \left\{ y^{\text{v}}+\varepsilon^{\text{v}},0 \right\},
\end{align}
implying that penalty term (\ref{delta v definition}) is convex on the ICNN input $x$. Similarly, if we substitute $y^{\text{ol}}$ and $y^{\text{rpf}}$ into equations (\ref{delta ol definition}) and (\ref{delta rpf definition}), $\delta^{\text{ol}}$ and $\delta^{\text{rpf}}$ are converted into convex functions of $x$.

\begin{figure}
    \centering
    \includegraphics[width=\linewidth]{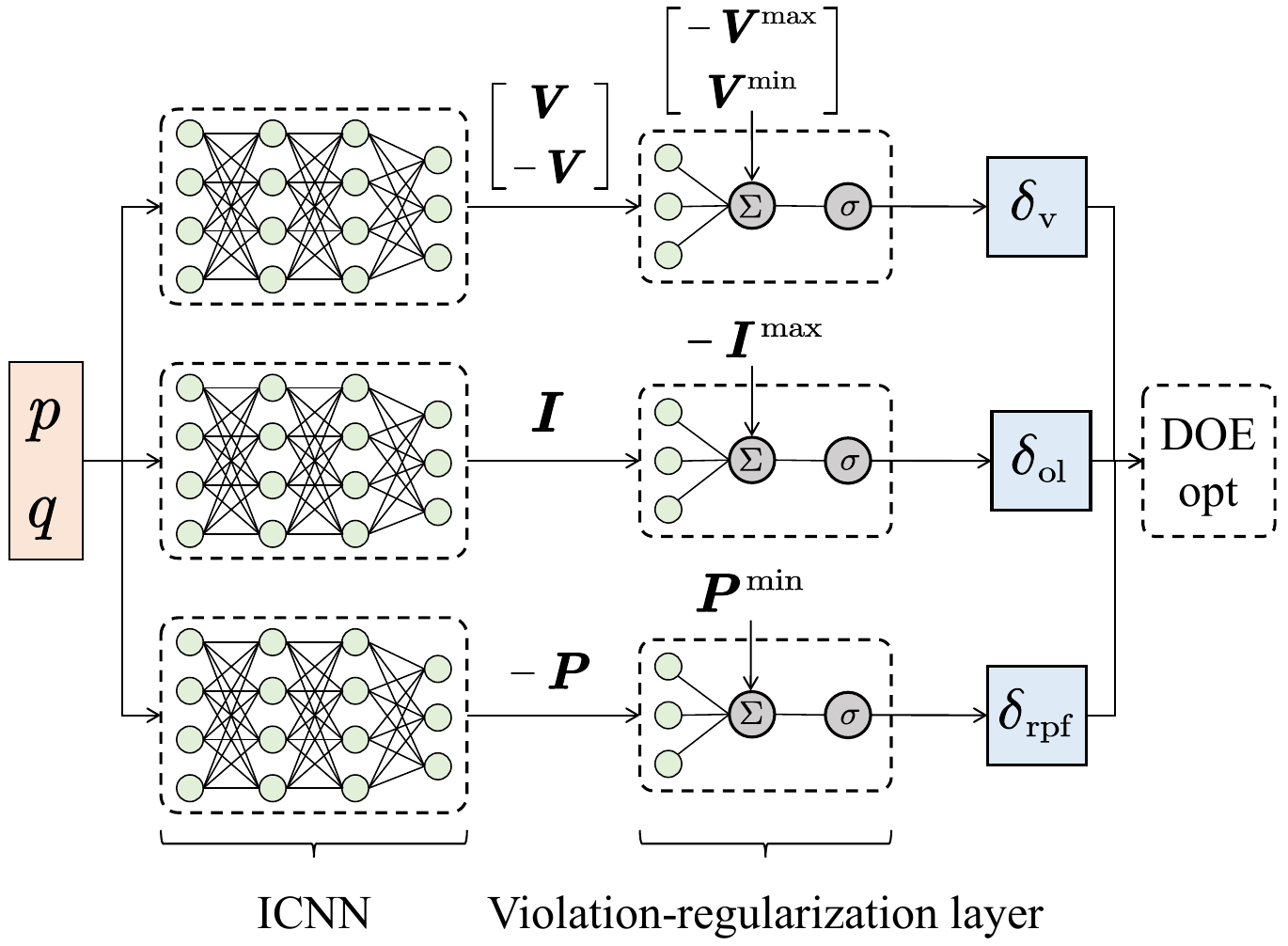}
    \caption{Proposed constraint embedding method that encodes constraint parameters in violation-regularization layers.}
    \label{constrain embedding}
\end{figure}
The proposed constraint embedding method is illustrated in Fig. \ref{constrain embedding}. By defining (\ref{y definition}) and (\ref{lim definition}), we encode the parameters of constraints (\ref{voltage limit}) - (\ref{reverse power flow limit}) in the violation-regularization layers, which converts the output of ICNNs to the penalty terms defined in equations (\ref{delta v definition}) - (\ref{delta rpf definition}). The ICNN models with violation-regularization layers can be written as:
\begin{subequations}
    \label{ICNN with regularization}
    \begin{align}
    &y^o=f(x|\theta^o), \\
    &\delta^o=\mathbf{1}^{\text{T}}\max \left( y^o+\varepsilon^{o},0 \right), o\in\{\text{v},\text{ol},\text{rpf}\}.
\end{align}
\end{subequations}

Directly plug the ICNN model\footnote{\textcolor{black}{The training of ICNN and the DOE optimization are separate processes. Once training is completed, the parameters of ICNNs (i.e., weights and biases) are fixed and used to parametrize the DOE optimization problem. With a training dataset that adequately covers the operating conditions, there is no need to retrain ICNN for DOE optimization across different time intervals.}} (\ref{ICNN with regularization}) into the DOE optimization problem (\ref{reformulated DOE optimization}) yields:
\begin{subequations}
\label{MIP DOE}
    \begin{align}
	\underset{p_{j,\phi}^{+}}{\text{min}}\;\;\;\;&w^{\text{DOE}}\sum_{j\in \mathcal{M},\phi\in\Phi}{\left| p_{j,\phi}^{\max}-p_{j,\phi}^{+} \right|}+w^{\text{loss}} P ^{\text{loss}}\notag\\
	&\quad \quad \quad \quad \quad \quad+w^{\text{v}}\delta ^{\text{v}}+w^{\text{ol}}\delta ^{\text{ol}}+w^{\text{rpf}}\delta^{\text{rpf}}\\
	\text{s}.\text{t}.\;\;\;\;&(\ref{DER power constraint}), (\ref{bus power definition}),\notag\\
    &x=\left[ p_{j,\phi};q_{j,\phi} \right] ,\\
    \label{nonlinear yo}
	&y^o=f(x|\theta^o),o\in \{\text{loss},\text{v},\text{ol},\text{rpf}\},\\
	&P^{\text{loss}}=y^{\text{loss}},\\
	&\delta^{o}=\mathbf{1}^{\text{T}}\max \left( y^{o}+\varepsilon^o,0 \right), \forall o\in \{\text{v},\text{ol},\text{rpf}\}.
\end{align}
\end{subequations}
Off-the-shelf solvers will interpret the optimization model (\ref{MIP DOE}) as an MILP and solve it with well-developed algorithms such as branch-and-bound. Compared to the original form of problem (\ref{reformulated DOE optimization}), which involves non-convex power flow constraints, an MILP problem is usually easier to solve. However, due to the integer variables, the solving could still be time-intensive, especially for large-scale systems. To further speed up the solving of problem (\ref{reformulated DOE optimization}), we leverage the exact inference feature of ICNN to relax the problem into a linear form. Note that the violation-regularization layers have the same structure as the hidden layers of ICNNs. The ICNN models with violation-regularization layers can be viewed as ICNNs with $K+1$ layer, where the non-negativity constraint on the weights of all $z^o_k$ holds:
\begin{align}
\label{non-negative}
    W^{z,o}_k\ge0, k=1, \cdots, K+1.
\end{align}
With the non-negativity constraint (\ref{non-negative}), we state and prove the tightness of the relaxation with the following theorem.
\begin{theorem}
    \label{ICNN DOE theorem}
    Given definitions (\ref{y definition}), (\ref{lim definition}) and constraint (\ref{non-negative}), optimizing the ICNN-based DOE optimization problem (\ref{MIP DOE}) is equivalent to optimizing the following LP problem:
\end{theorem}
\begin{subequations}
\label{linear DOE}
    \begin{align}
	\underset{p_{j,\phi}^{+}}{\text{min}}\;\;\;\;&w_{\text{DOE}}\sum_{j\in \mathcal{M},\phi\in\Phi}{\left| p_{j,\phi}^{\max}-p_{j,\phi}^{+} \right|}+w^{\text{loss}}\delta^{\text{loss}}\notag\\
	&\quad \quad \quad \quad \quad \quad+w^{\text{v}}\delta ^{\text{v}}+w^{\text{ol}}\delta ^{\text{ol}}+w^{\text{rpf}}\delta ^{\text{rpf}}\\
	\text{s}.\text{t}.\;\;\;\;&(\ref{DER power constraint}), (\ref{bus power definition}),\notag\\
    &x=\left[ p_{j,\phi};q_{j,\phi} \right] ,\\
    \label{linear inequality z1o}
    &z_{1}^{o}\ge W_{0}^{x,o}x+b_{0}^{o},z_{1}^{o}\ge 0,\notag\\
    &\quad\quad\quad\quad\quad\quad\quad\quad\quad\quad\; \forall o\in \{\text{loss},\text{v},\text{ol},\text{rpf}\},\\
    \label{linear inequality zko}
    &z_{k+1}^{o}\ge W_{k}^{z,o}z_{k}^{o}+W_{k}^{x,o}x+b_{k}^{o}, z_{k+1}^{o}\ge 0, \notag\\
    &\quad\quad\; k=1,\cdots,K-1,\; \forall o\in \{\text{loss},\text{v},\text{ol},\text{rpf}\}, \\
    \label{nu}
    &\nu^o\ge W_{K}^{z,o}z_{K}^{o}+W_{K}^{x,o}x+b_{K}^{o}+\varepsilon^o,\nu^o\ge 0, \notag\\
    &\quad\quad\quad\quad\quad\; \delta^o\ge \mathbf{1}^{\text{T}}\nu^o,\; \forall o\in \{{\text{loss}},\text{v},\text{ol},\text{rpf}\},
\end{align}
\end{subequations}
where $\delta^{\text{loss}}$, $\nu^{\text{v}}$, $\nu^{\text{ol}}$, $\nu^{\text{rpf}}$ are auxiliary variables and $\varepsilon^{\text{loss}}=0$.
\begin{proof}
The proof of Theorem~\ref{ICNN DOE theorem} is derived from the exact inference of ReLU-activated ICNN. Although there are four ICNNs that handle different terms in the objective function, they work independently. Thus, we first prove the equivalence for one ICNN, and the results can be extended to all four ICNNs. Note that weighing factors $w^{\text{loss}}$, $w^{\text{v}}$, $w^{\text{ol}}$ and $w^{\text{rpf}}$ are non-negative. With a non-zero weight $w^o$, solving problem (\ref{linear DOE}) essentially minimizes the corresponding term $\delta^o$. Since inequality constraints (\ref{nu}) are convex, minimizing $\delta^o$ implies that these constraints must be tight:
\begin{align}
    &\delta^o = \mathbf{1}^{\text{T}}\nu^o, \\
    &\nu^o = W_{K}^{z,o}z_{K}^{o}+W_{K}^{x,o}x+b_{K}^{o}+\varepsilon^o \quad \text{or} \quad \nu^o= 0.
\end{align}
When $\nu^o \ne 0$, minimizing $\nu^o$ is equivalent to minimizing $z_{K}^{o}$, since constraint (\ref{non-negative}) enforces the weight matrix $W_{K}^{z,o}$ to be non-negative. Assuming that we have a given input $x$, minimizing $\delta^o$ requires minimizing $W_{K}^{z,o}$ subject to constraints (\ref{linear inequality z1o}) and (\ref{linear inequality zko}), which coincide with the exact inference of ICNN described in problem (\ref{exact inference of ICNN}). Thus, linear relaxations (\ref{linear inequality z1o}) and (\ref{linear inequality zko}) exactly recover the nonlinearity of ICNN models (\ref{nonlinear yo}), without introducing integer variables. The equivalence between problem (\ref{MIP DOE}) and problem (\ref{linear DOE}) can be easily extended from one ICNN model to multiple ICNN models and from a given $x$ to all possible $x$ that satisfy constraints (\ref{DER power constraint}) and (\ref{bus power definition}). This completes the proof of Theorem~\ref{ICNN DOE theorem}.
\end{proof}

Theorem~\ref{ICNN DOE theorem} indicates that, by exploiting the convexity and linearity of ICNNs, instead of solving the non-convex DOE optimization problem (\ref{reformulated DOE optimization}), we can equivalently optimize an LP problem (\ref{linear DOE}). Although ICNN models introduce auxiliary variables to the optimization problem, the linearity of the LP problem (\ref{linear DOE}) allows it to be solved in a very efficient manner by off-the-shelf solvers. 

\textcolor{black}{Compared to existing works in the literature, this paper is the first to leverage neural networks to directly replicate the power flow mappings from nodal power to nodal voltage, line current, and active power flow. Compared to other ICNN-assisted optimization methods, such as Ref.~\cite{wu_transient_2024}, the proposed constraint embedding method and input-output design provide a rigorous way to integrate ICNNs into optimization problems, where the outputs of ICNNs are bounded both above and below, as proved in Theorem~\ref{ICNN DOE theorem}. In contrast to existing neural-network-based optimization techniques, such as those in Refs.~\cite{wu_transient_2024,chen_scheduling_2021}, the proposed method can adapt flexibly to changes in constraint parameters without the need for retraining. This is achieved by encoding the constraint parameters within the violation-regularization layers of ICNNs.} As we discussed in the previous section, utilities need to update DOEs for DERs frequently in response to the latest DER power forecasts and grid operating conditions. The improvement in optimization efficiency and parameter adaptability will significantly reduce the computation time of optimal DOE, especially in large-scale systems with substantial DERs.

\subsection{Constraint Retrenchment Strategy}
\textcolor{black}{In the discussions above, we focus on improving the tractability of the DOE optimization problem from the perspective of operations research. Although LP problems can be efficiently solved, longer solution time is required when the number of variables and constraints grows larger. Intuitively, the solution time of the relaxed DOE optimization problem (\ref{linear DOE}) is positively correlated with the number of neurons used in ICNNs. Thus, to preserve solution efficiency in large-scale systems, we want to keep the number of neurons to its minimum.} To this end, we propose a constraint retrenchment strategy that leverages the radial structure of distribution system topologies and prunes out unbinding constraints. With the unbinding constraint set, we can remove some entries of the output vectors $y^o$ and thereby reduce the dimensions of ICNN output layers.

\begin{figure}
    \centering
    \includegraphics[width=0.8\linewidth]{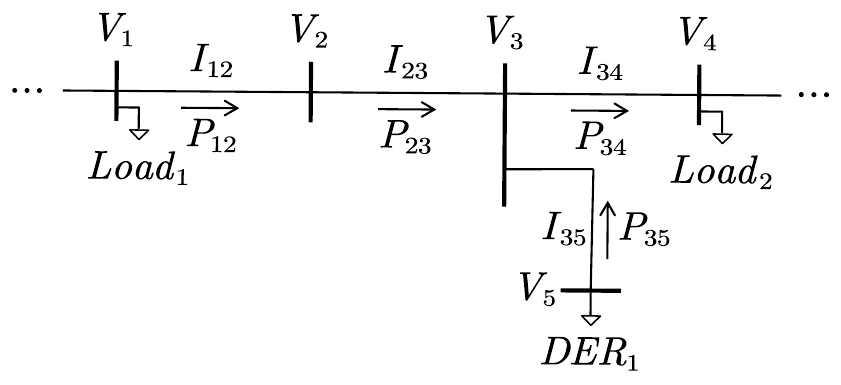}
    \caption{Illustration of a radial distribution system.}
    \label{constraint retrenchement}
\end{figure}
We use Fig.~\ref{constraint retrenchement}, which illustrates a part of a radial distribution system with five nodes, to discuss the principle of constraint retrenchment. We define a subset of nodes $\bar{\mathcal{M}}\subset\mathcal{M}$, which contains the nodes that only transit power and have no load or DER connected. The set of other nodes is defined as $\Tilde{\mathcal{M}}=\mathcal{M}-\bar{\mathcal{M}}$. For the buses in set $\bar{\mathcal{M}}$, e.g., node 2 and node 3 in Fig.~\ref{constraint retrenchement}, their voltage magnitudes are bounded by the voltage magnitudes of the adjacent nodes, i.e., node 1, node 4, and node 5. Assume that we set the same upper and lower voltage limits for all buses. If the voltage magnitudes of nodes in $\Tilde{\mathcal{M}}$ are within the limits in constraint (\ref{voltage limit}), the voltage limits of nodes in $\bar{\mathcal{M}}$ are not breached. Thus, for DOE optimization, we can let ICNN learn the voltage magnitudes of nodes in $\Tilde{\mathcal{M}}$, instead of all nodes in $\mathcal{M}$. Similarly, if the same thermal capacity is posed to all lines, the current magnitude of line (2,3) is unlikely to reach the limit before violation of the thermal capacities of line (1,2), line (3,4), and line (3,5). Therefore, we can let ICNN focus on the current magnitudes of the lines that are connected to a node in $\Tilde{\mathcal{M}}$, instead of all lines. Finally, since reverse power flow only happens when DERs are injecting power into the grid, we can let ICNN prioritize the active power of the immediate upstream lines of DER nodes. After retrenching, ICNNs learn the mapping between fewer variables, reducing the number of neurons in ICNN and the number of variables in the equivalent problem (\ref{linear DOE}). Thus, computation burdens can be reduced.

\textcolor{black}{The proposed constraint retrenchment strategy reduces the computational burdens of DOE optimization by reducing the dimensionality of the reformulated problem. Also, with fewer output dimensions, the computational burden in ICNN training is reduced. Thus, the constraint retrenchment strategy complements the proposed constraint embedding method, which accelerates the optimization by improving the tractability of the problem.}
\begin{remark}
    \textcolor{black}{It is worth noting that the constraint retrenchment is derived using a radial network. Although most existing distribution systems are operated with a radial topology, meshed topology and loops could also exist in some scenarios. When the topology is not radial, the DOE optimization problem becomes more complicated. The constraint retrenchment strategy remains applicable to the nodes that only transit power and have no load or DER connection. Furthermore, in meshed or looped topology, it is possible to sectionize the system into partitions, where some of the partitions have radial structure. Then, the proposed constraint retrenchment strategy can be applied to those partitions. In practical applications, conducting case-dependent analyses and data-based verification is essential to ensure system integrity is not compromised.}
\end{remark}

\section{Case Study}
\label{section 4}
In this section, the effectiveness of the proposed method is validated using a modified IEEE 123-node test feeder and a modified EPRI Ckt5 test feeder, which are both three-phase unbalanced distribution systems. First, we demonstrate the accuracy of ICNN in learning the mapping between physical quantities in the distribution system. Then, the performance of the ICNN-based DOE optimization method is compared with benchmark methods to demonstrate the advantage of the proposed method.

\subsection{Experiment Setup}
\subsubsection{Test systems}
\begin{figure}
    \centering
    \includegraphics[width=\linewidth]{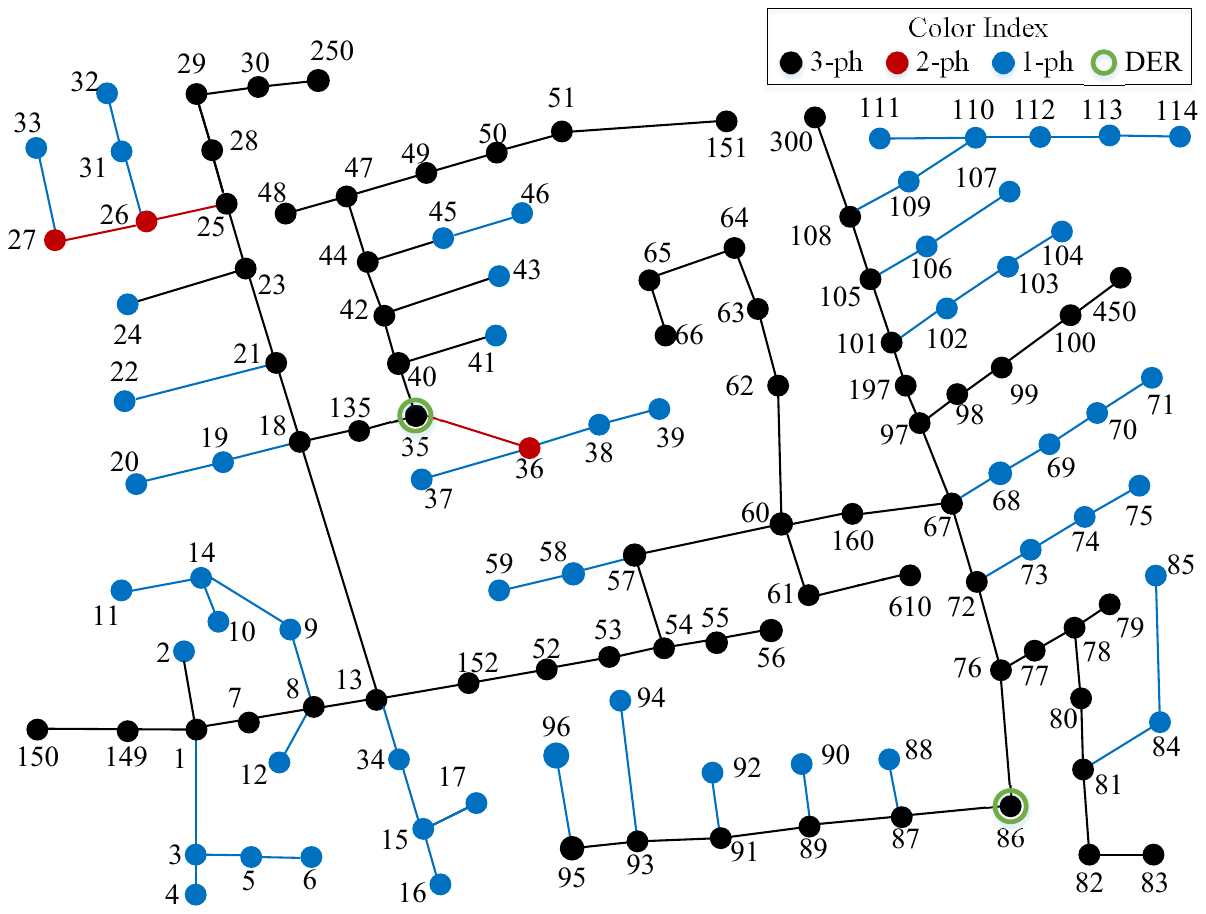}
    \caption{Modified IEEE 123-bus test feeder.}
    \label{123 node}
\end{figure}

\begin{figure}
    \centering
    \includegraphics[width=\linewidth]{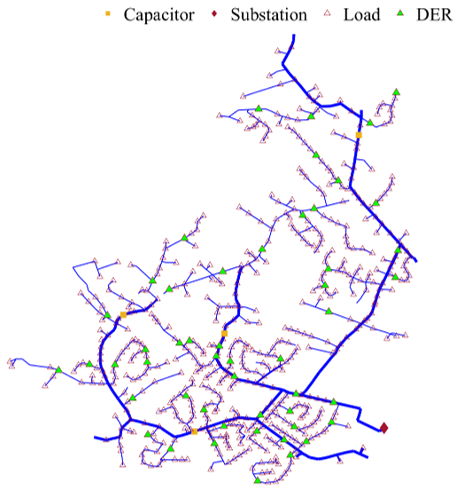}
    \caption{\textcolor{black}{Modified EPRI Ckt5 test feeder.}}
    \label{ckt5}
\end{figure}

Two test systems are developed based on the IEEE 123-node test feeder and the EPRI Ckt5 test feeder, as introduced below:
\begin{itemize}
    \item The IEEE 123-bus test feeder is a radial distribution system at a nominal voltage of 4.16 kV, featuring unbalanced loading. In the system, there are three-phase, two-phase, and single-phase lines, making the power flow model more complex. To clearly demonstrate the performance of different benchmark methods, we install two DERs in the test system. Table~\ref{DER configuration} outlines the locations, upper/lower active power limits, and reactive power setpoints of individual DERs. The system topology and DER locations are visualized in Fig~\ref{123 node}. We assume the following limits for the modified IEEE 123-node test feeder: $v^{\min}=0.9 \text{ p.u.}$, $v^{\max}=1.1 \text{ p.u.}$, $I^{\max}=500 \text{ A}$ and $P^{\min}=-125 \text{ kW}$.
    \item \textcolor{black}{The EPRI Ckt5 test feeder is also a radial distribution system, with a nominal voltage of 12.47 kV. There are 1039 nodes, connected by three-phase, two-phase, and single-phase lines. In this paper, we focus on the primary side of the distribution system and aggregate the secondary network to the distribution transformer level, resulting in 591 single-phase loads. On top of these loads, 50 single-phase PVs are installed to validate the scalability of the proposed methods. The topology and the DER locations are illustrated in Fig.~\ref{ckt5}. The voltage limits are set to $v^{\min}=0.9 \text{ p.u.}$ and $v^{\max}=1.1 \text{ p.u.}$, respectively. The overloading limits and reverse power flow limits are defined according to the line code provided by the original system.}
\end{itemize}
\textcolor{black}{In the experiments, we set the weighting factors $w^{\text{v}}$, $w^{\text{ol}}$ and $w^{\text{rpf}}$ to large values, such that the optimization results coincide with the original DOE optimization problem.} OpenDSS~\cite{dugan_open_2010}, whose power flow solutions are considered accurate, is utilized to validate the results. We also use the OpenDSS model to generate 30,000 power flow snapshots for ICNNs to learn, by randomly sampling the active and reactive power of the buses. \textcolor{black}{For each snapshot, a random load level is first selected, ranging from zero to 1.2 times the aggregated spot load provided by the original test case. A local generation level is then randomly selected from $[0,1.1]$ times of the load level. Then the loads and local generations are randomly distributed across the nodes. In this way, we simulate operation scenarios with both light/heavy loading levels and low/high DER penetration levels. Note that the number of DERs and their locations are not fixed when generating training data. Thus, the trained ICNNs learn the power flow equations, without assuming specific DER configurations, enabling them to generalize effectively under varying load profiles and DER settings, without the need for retraining. The proposed ICNNs demonstrate satisfactory accuracy in the dataset, as shown in the next subsection.} The active and reactive load profiles are synthesized based on smart meter data in the Midwest U.S. The power profiles of DERs are synthesized from solar farms and electric vehicle charging stations in the Midwest U.S.

\begin{remark}
    \textcolor{black}{In this paper, we sample the operation data without assuming the nominal operating conditions. In this way, the ICNNs exhibit decent performance across a wide range of operating conditions. However, if the distribution system enters an extreme operating condition that falls outside the sampled range of operating conditions, the performance of ICNNs could deteriorate. If the utility has historical operational data\footnote{\textcolor{black}{Real-world operational data could have data quality issues, such as noise. Usually, the impact of noise can be minimized by proper pre-processing, such as filtering. If noisy data is used to train ICNNs, the accuracy performance will deteriorate. Since data quality issues are not the main focus of this paper, we assume that the data used for training is clean and precise.}}, it can leverage this data to identify common operating conditions and build the training set accordingly. The identified common operating conditions can also be used to detect extreme operating conditions. If an extreme condition is detected, the utility should exercise caution when using ICNNs and run model-based power flow simulations to verify and ensure the feasibility of the optimization results.}
\end{remark}

\begin{table}[bt]
\centering
\caption{DER Configurations.}
\label{DER configuration}
\begin{tabular}{C{1}C{0.6}C{1.6}C{1.6}C{1.6}}
\toprule
DER \#  & Bus & $p_j^{\text{max}}$ (kW) & $p_j^{\text{min}}$ (kW) & $q_j$ (kVar) \\ \midrule
1 & 35   & 800                    & 0                       & 80          \\
2 & 86  & 500                     & -500                    & 0        
\\ \bottomrule
\end{tabular}
\end{table}
\begin{table}[bt]
\centering
\caption{Hidden Layer Configuration and NMAE of Neural Networks.}
\label{ICNN config table}
\begin{tabular}{C{0.8}C{3}C{1.5}C{1.5}}
\toprule
\multirow{2}{*}{} & \multirow{2}{*}{Hidden layer configuration} & \multicolumn{2}{c}{NMAE} \\ \cmidrule(l){3-4} 
                  &                    & ICNN         & MLP         \\ \midrule
$y^{\text{loss}}$ & [128,64]          & 0.00032 & 0.00021\\
$y^{\text{v}}$    & [256,256,128,128] & 0.00040 & 0.00033\\
$y^{\text{ol}}$   & [256,256,128,64]  & 0.0011  & 0.00076\\
$y^{\text{rpf}}$  & [128,64]          & 0.00019 & 0.00020\\ 
\bottomrule
\end{tabular}
\end{table}
\subsubsection{Benchmarks:}
To validate the effectiveness of the ICNN-based method, we compare it with two benchmark methods:
\begin{itemize}
    \item \textbf{B0:} DER power takes the upper limit for $p_{j,\phi}^+$ and lower limit for $p_{j,\phi}^-$, without considering DOE.
    \item \textbf{B1:} Proposed ICNN-aided DOE optimization method with linear relaxation, as shown in problem (\ref{linear DOE}).
    \item \textbf{B2:} Proposed ICNN-aided DOE optimization method without linear relaxation, as shown in problem (\ref{MIP DOE}). \textcolor{black}{This method yields the same solution as B2. It is included to demonstrate the tightness of the proposed linear relaxation and the resulting improvement of solution efficiency.}
    \item \textbf{B3:} Adopt linear DistFlow (LinDistFlow)~\cite{gan_convex_2014} model to replace power flow constraints and optimize DOE in problem (\ref{reformulated DOE optimization}). The formulation is provided in Appendix~\ref{LinDistFlow formulation}. \textcolor{black}{This benchmark represents linear-relaxation-based methods that are widely adopted by the industry to handle non-convex power flow.}\footnote{\textcolor{black}{Nonlinear model-based power flow relaxations, such as second-order cone programming (SOCP) relaxation, have been widely applied in the literature. In the field of DOE optimization, the inexactness of SOCP relaxation has been pointed out be researchers in Ref.~\cite{moring_Inexactness_2023}. This is because the tightness of the SOCP relaxation relies on specific conditions, such as monotone objective functions, no reverse power congestion, and no binding lower voltage limits, which usually do not hold in the context of DOE optimization. Otherwise, SOCP relaxation could be an effective approach for addressing non-convex power flow equations.}}
    \item \textbf{B4:} Use ReLU-activated MLPs that share the same hidden layer configurations and input-output with the ICNNs. The DOE optimization problem with MLPs is reformulated into MILP problems, inspired by Ref.~\cite{chen_efficient_2024}. \textcolor{black}{This benchmark represents the state-of-the-art neural-network-based replication of power flow equations, which can be embedded into subsequent optimization problems.}
\end{itemize}
The above benchmark methods are implemented using a PC with an Intel(R) Xeon(R) W-1370 CPU and 32 GB RAM. Gurobi is used as the optimization solver~\cite{gurobi_optimization_gurobi_2022}.

\subsection{Accuracy of ICNN}
\label{section 4b}
For benchmarks B1, B2, and B4, we train ICNNs and MLPs to learn the mapping from bus power to active power loss, bus voltage magnitude, line current magnitude, and line active power. The configurations of neural networks and the normalized mean absolute error (NMAE) are outlined in Table~\ref{ICNN config table}.

The metric NMAE is defined as the mean absolute error divided by a normalization factor:
\begin{align}
    \text{NMAE}_y=\frac{\frac{1}{n}\sum_{i=1}^n{\left| \hat{y}_i-y_{i} \right|}}{\tilde{y}},
\end{align}
where $n$ is the sample number of testing data; $\hat{y}_i$ and $y_{i}$ are the ICNN output value and actual value of the $i$th sample; $\tilde{y}$ is the normalization factor. In this paper, we use the nominal voltage, substation transformer rated power, and line rated current as the normalization factors for voltage, power, and current, respectively. As shown in Table~\ref{ICNN config table}, although MLPs' NMAE is slightly lower, ICNNs demonstrate competitive and satisfactory performance in learning the desired mappings.

\begin{figure}
    \centering
    \subfloat[Active power loss.\label{loss forecast}]{%
       \includegraphics[width=0.49\linewidth]{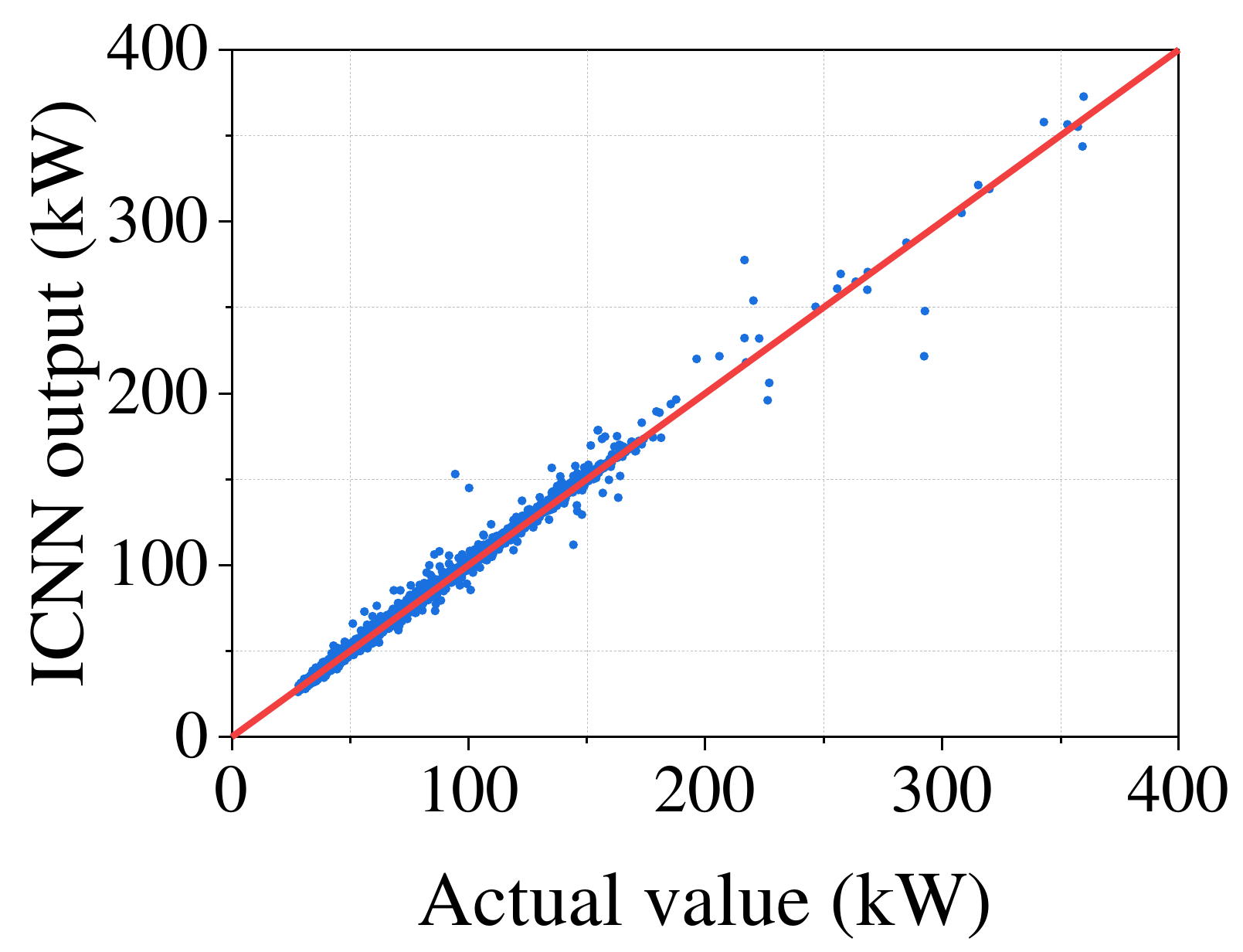}}
       \hfill
    \subfloat[Voltage magnitude.\label{voltage forecast}]{%
       \includegraphics[width=0.49\linewidth]{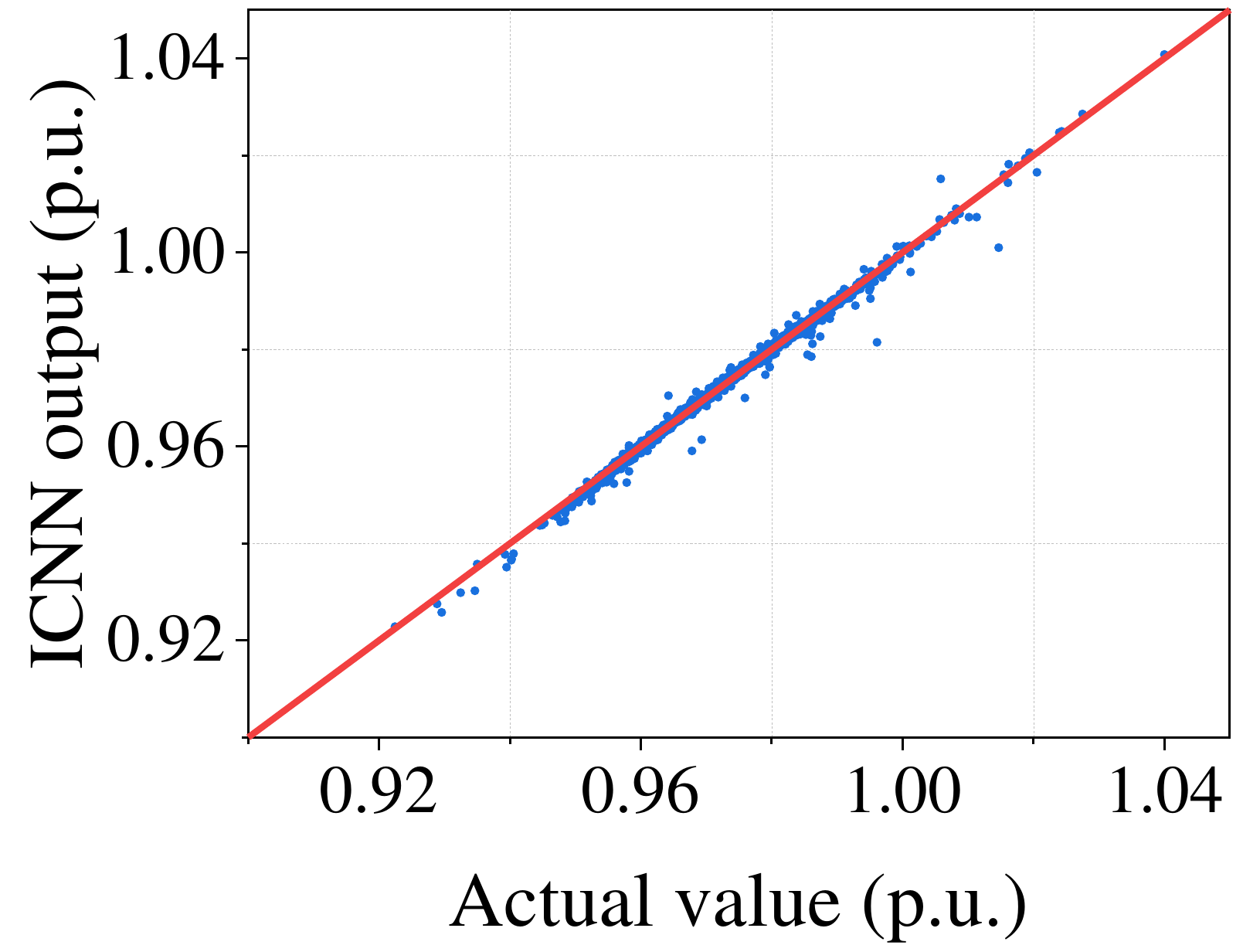}}\\
       \subfloat[Line current magnitude.\label{current forecast}]{%
       \includegraphics[width=0.49\linewidth]{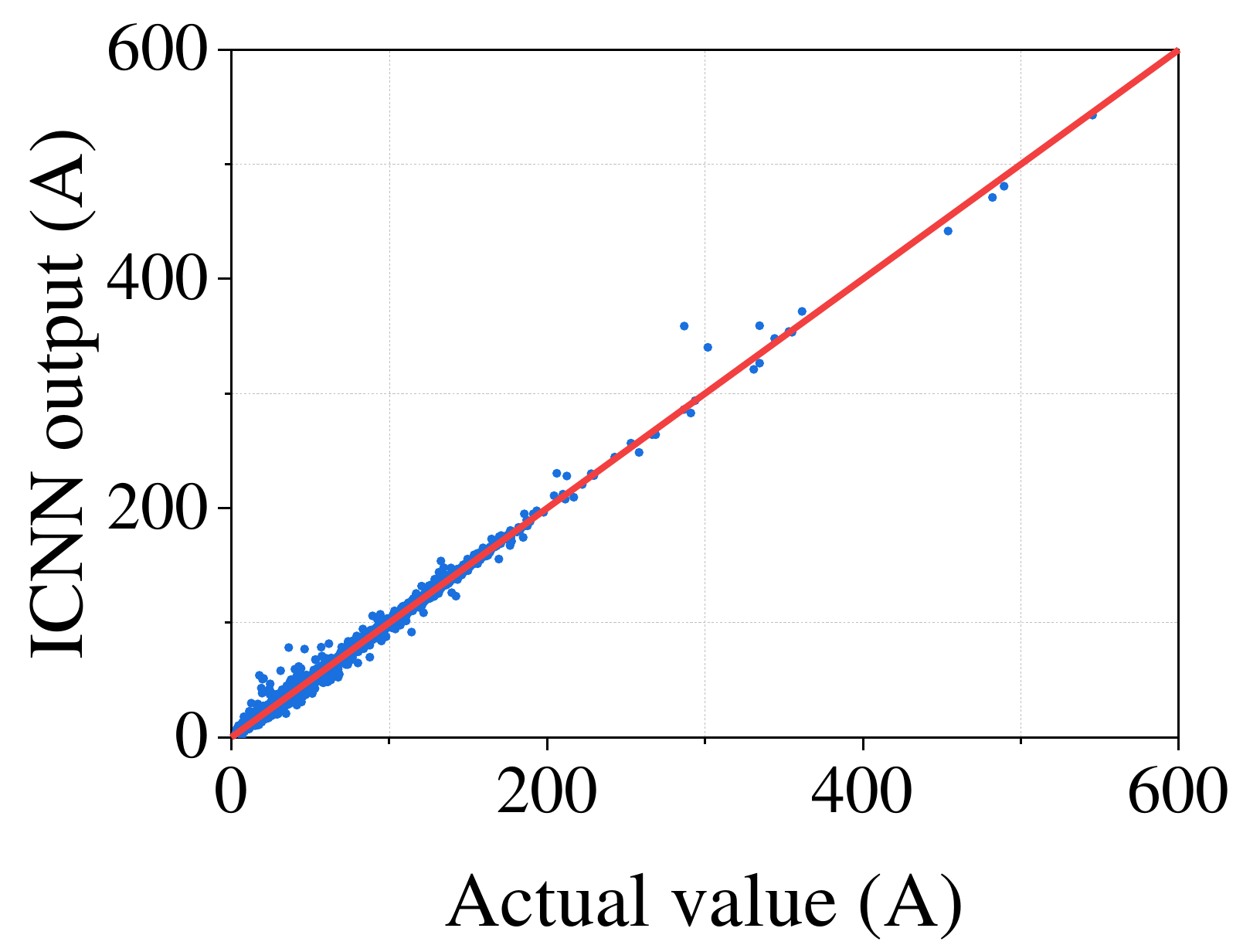}}
       \hfill
    \subfloat[Line active power.\label{power forecast}]{%
       \includegraphics[width=0.49\linewidth]{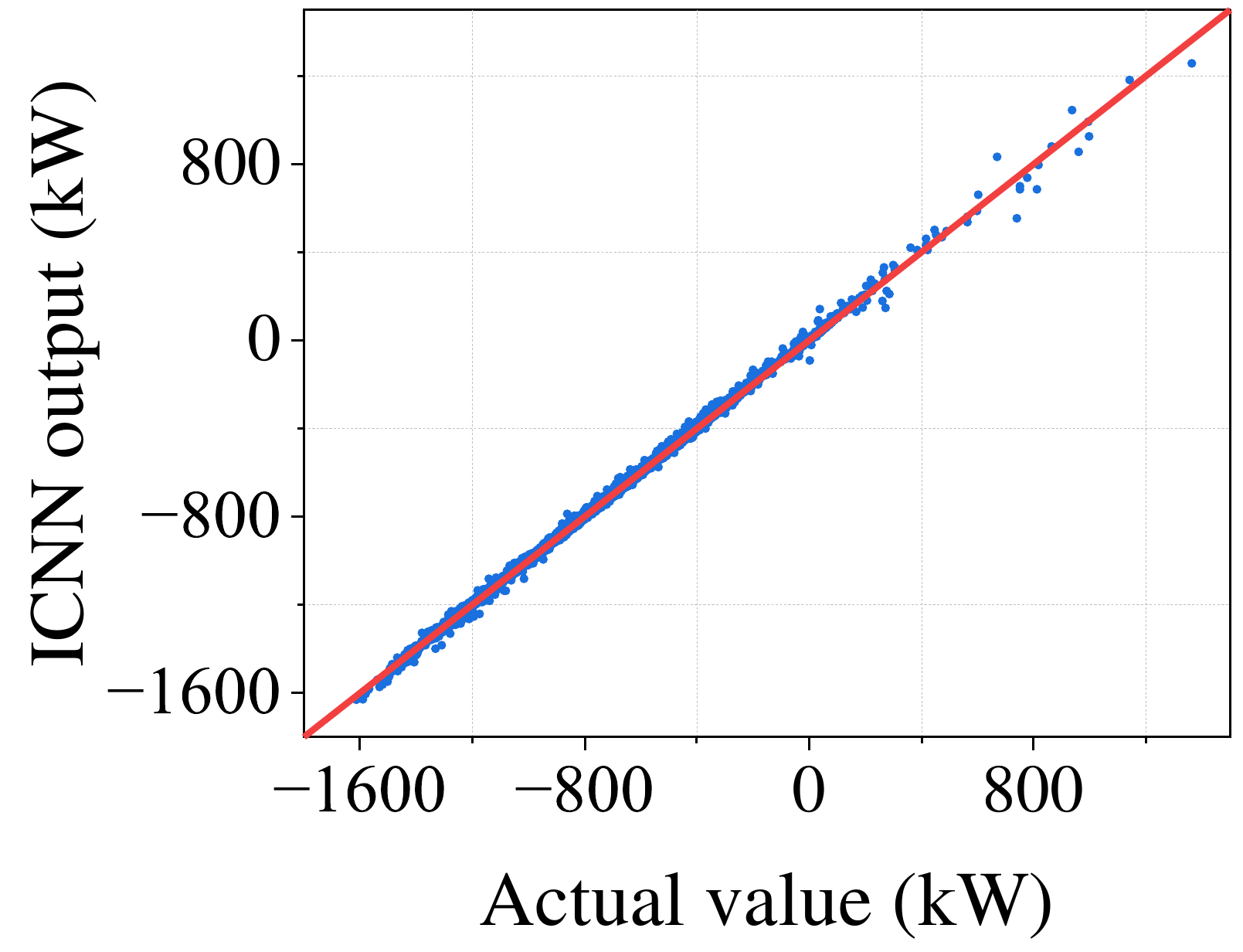}}\\
    \caption{Comparison of ICNN outputs with actual values of IEEE 123-node test feeder.}
    \label{ICNN accuracy}
\end{figure}

\begin{figure}
    \centering
    \subfloat[Active power loss.\label{loss ckt5}]{%
       \includegraphics[width=0.49\linewidth]{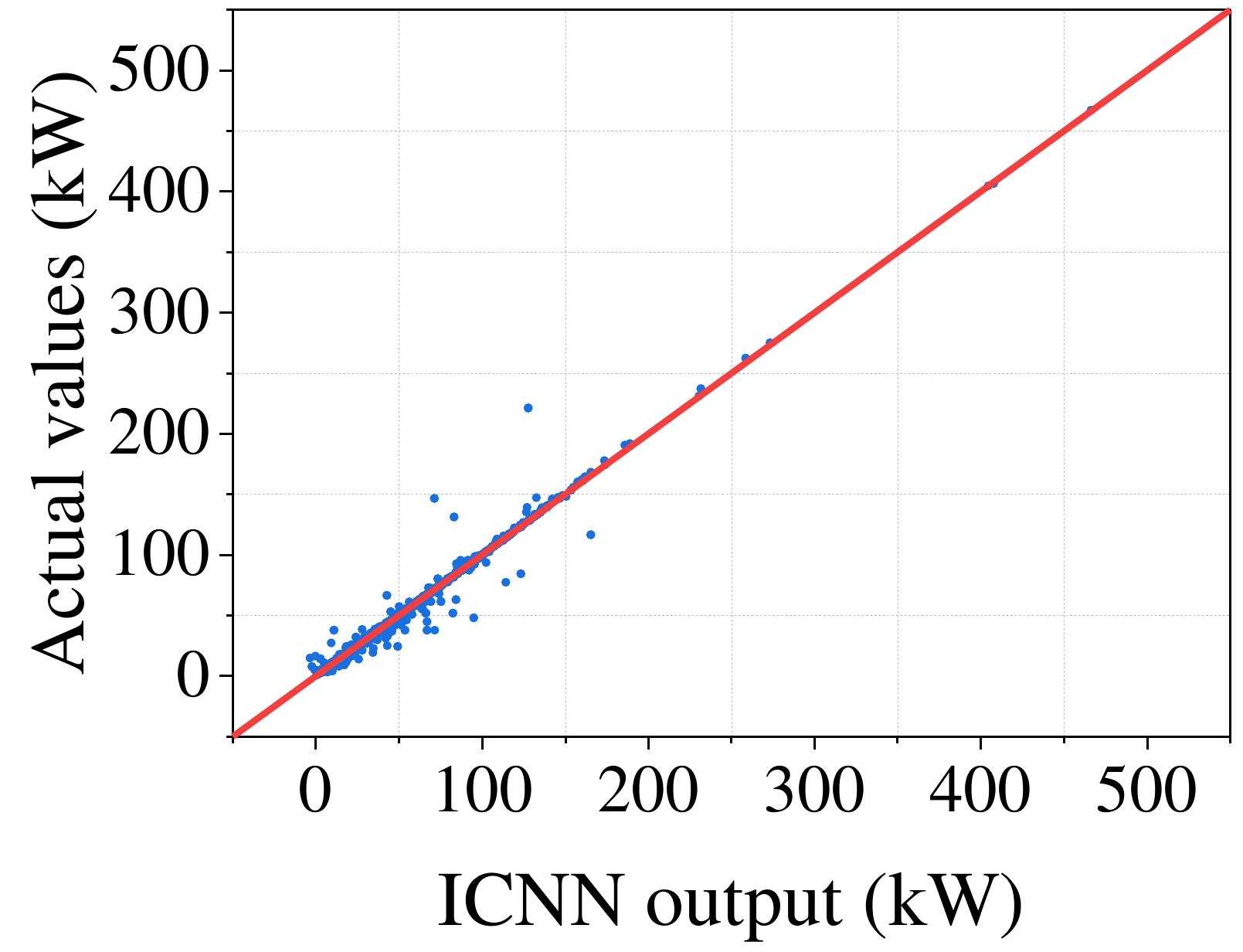}}
       \hfill
    \subfloat[Voltage magnitude.\label{voltage ckt5}]{%
       \includegraphics[width=0.49\linewidth]{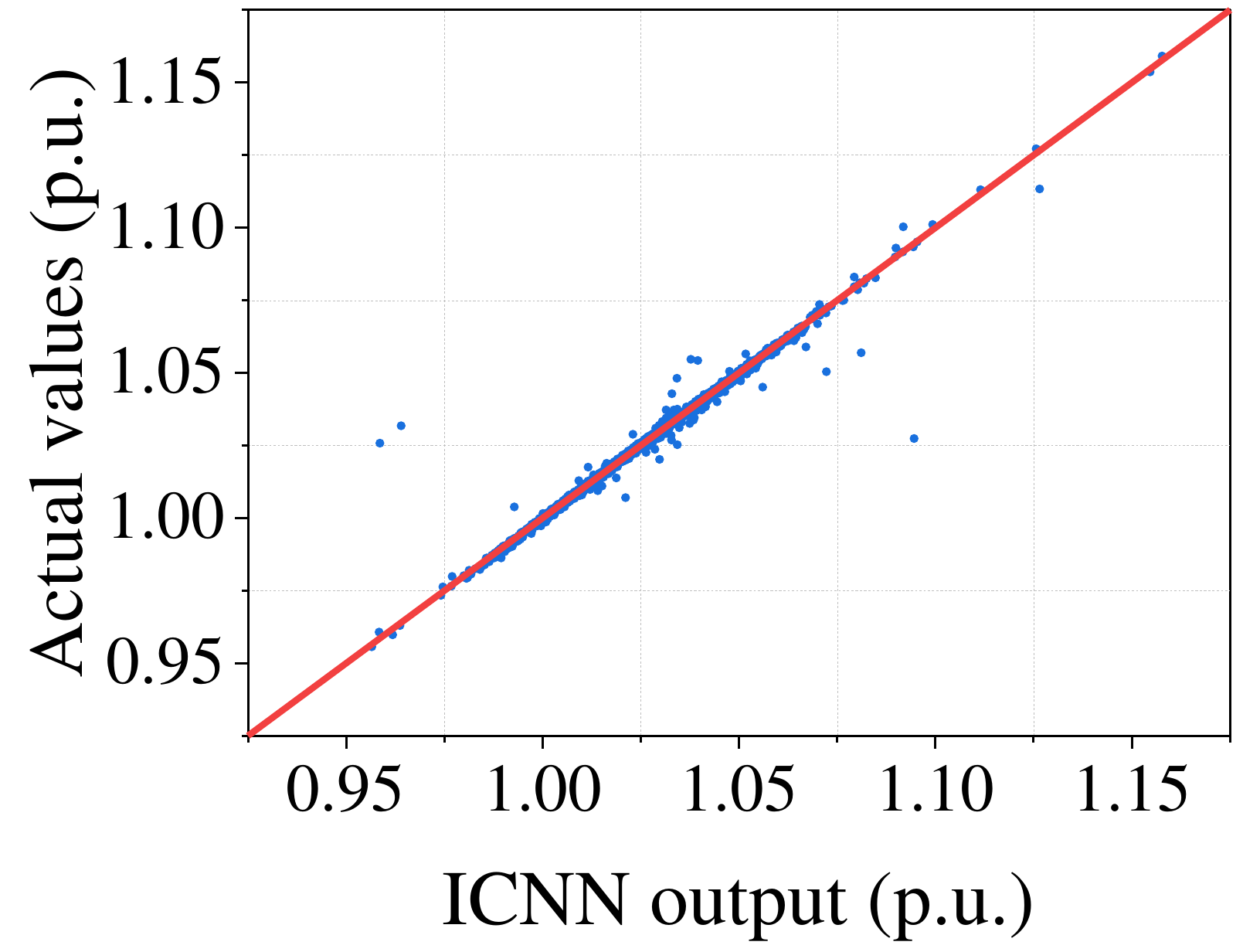}}\\
       \subfloat[Line current magnitude.\label{current ckt5}]{%
       \includegraphics[width=0.49\linewidth]{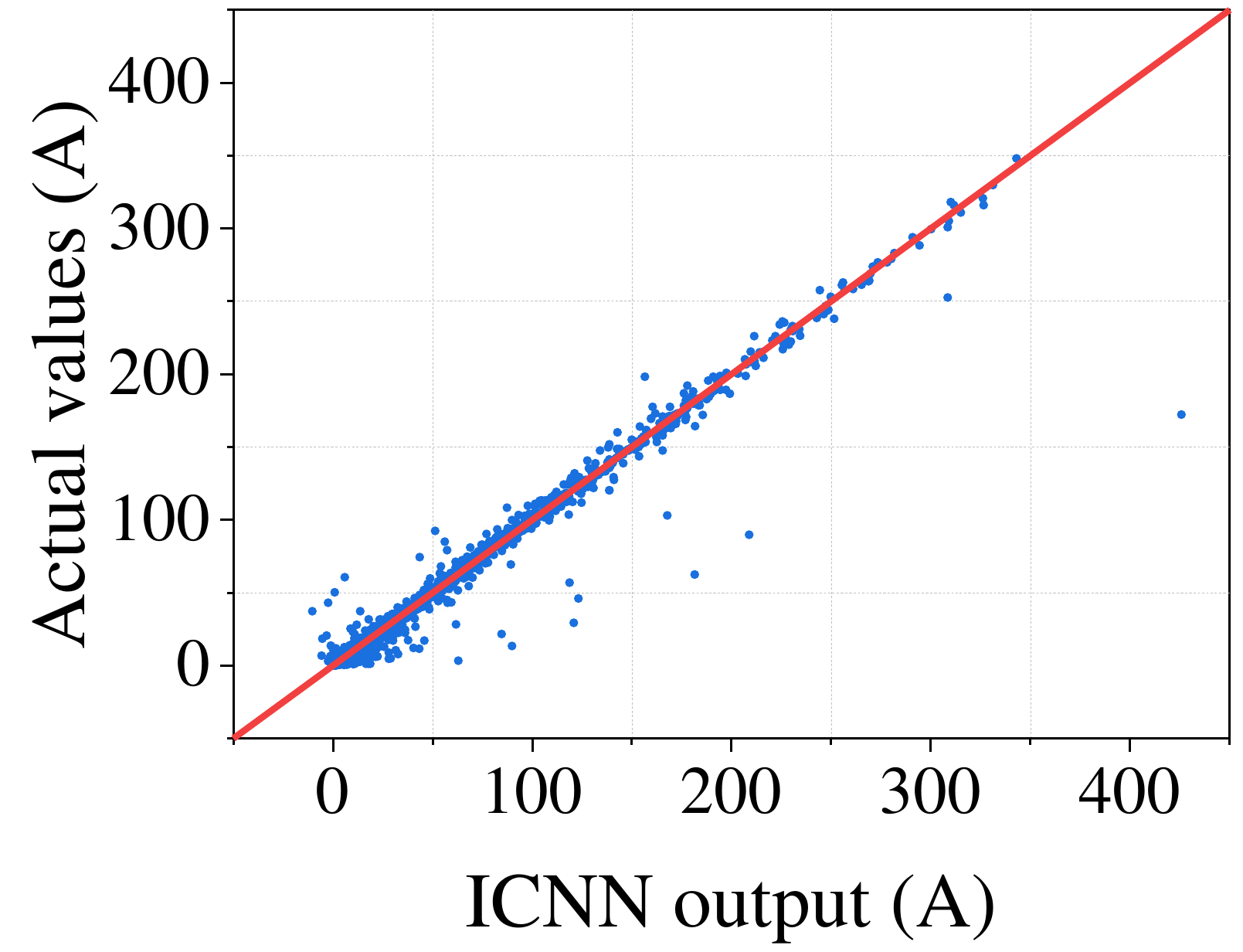}}
       \hfill
    \subfloat[Line active power.\label{power ckt5}]{%
       \includegraphics[width=0.49\linewidth]{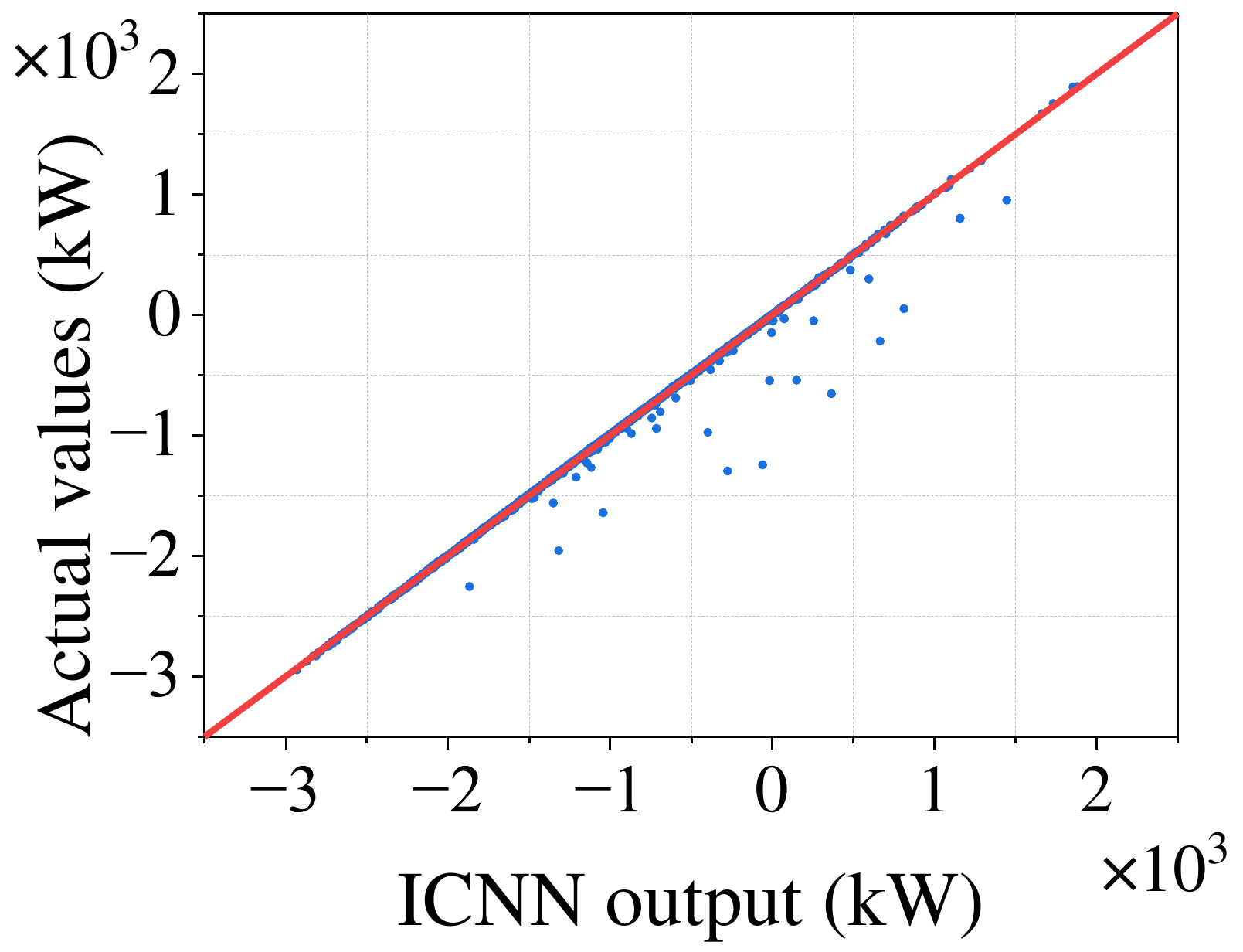}}\\
    \caption{\textcolor{black}{Comparison of ICNN outputs with actual values of EPRI Ckt5 test feeder.}}
    \label{ICNN accuracy ckt5}
\end{figure}

To further illustrate the accuracy of the ICNNs, we visualize the ICNN outputs versus the actual values in Figs.~\ref{ICNN accuracy} and \ref{ICNN accuracy ckt5}. The blue dots represent data points, where the actual values serve as x-values and the ICNN output as y-values. The red lines are reference lines with slopes equal to one. \textcolor{black}{The blue dots are concentrated around the red lines, implying that for most of the data points in the testing data, the ICNN outputs are very close to the actual values, which further indicates ICNNs' capability in learning and replicating the power flow equations.} \textcolor{black}{Since topological information is not part of the input of ICNNs, they might not generalize well under changes in network topology or line impedance. Possible approaches to address this issue include utilizing network parameters and network topology as input features and employing graph neural network structures. Nonetheless, this aspect is not the primary focus of this paper, and we highlight it as an important direction for future work.}


\begin{figure}
    \centering
    \subfloat[Active power loss.\label{nmae loss}]{%
       \includegraphics[width=0.49\linewidth]{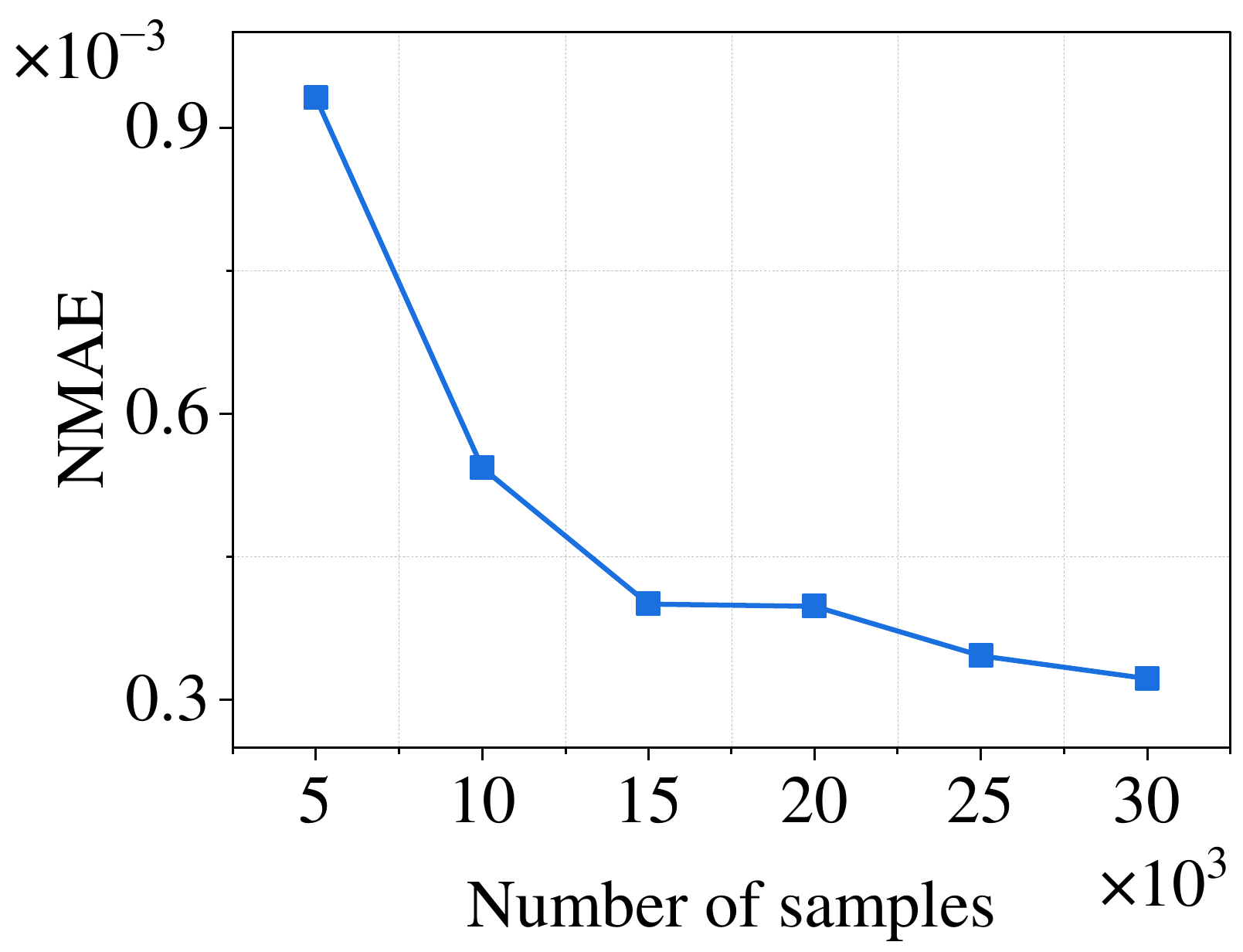}}
       \hfill
    \subfloat[Voltage magnitude.\label{nmae voltage}]{%
       \includegraphics[width=0.49\linewidth]{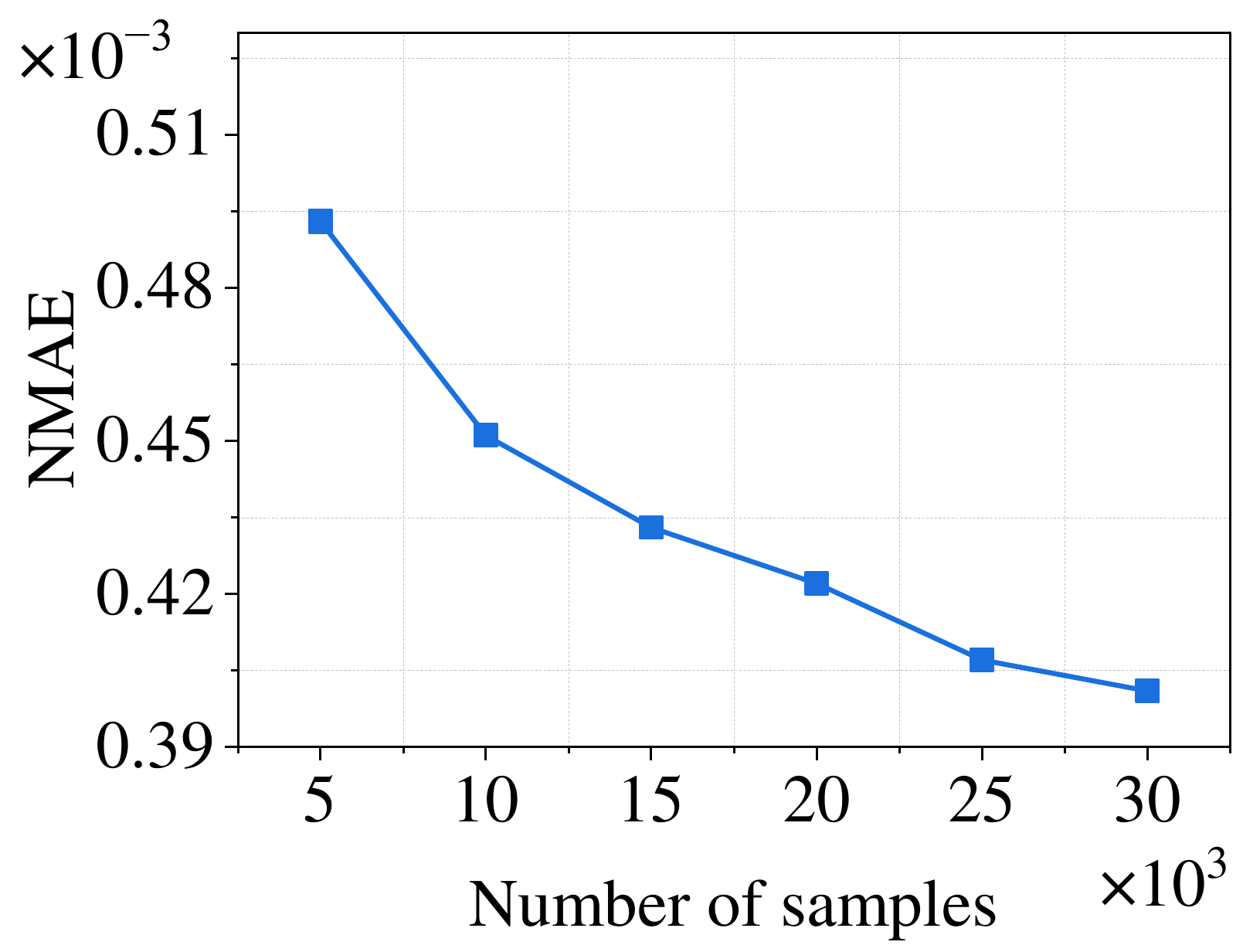}}\\
       \subfloat[Line current magnitude.\label{nmae current}]{%
       \includegraphics[width=0.49\linewidth]{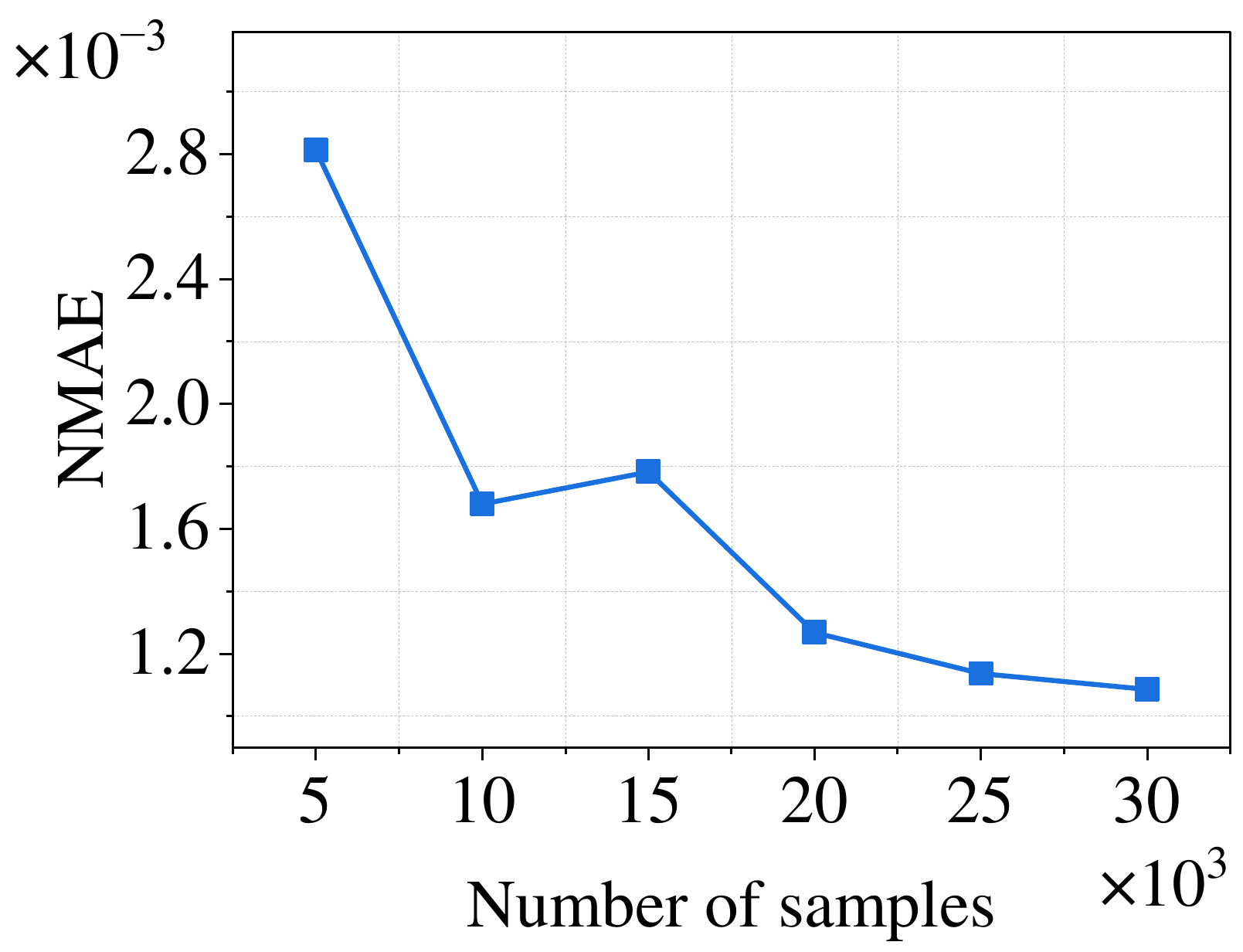}}
       \hfill
    \subfloat[Line active power.\label{nmae power}]{%
       \includegraphics[width=0.49\linewidth]{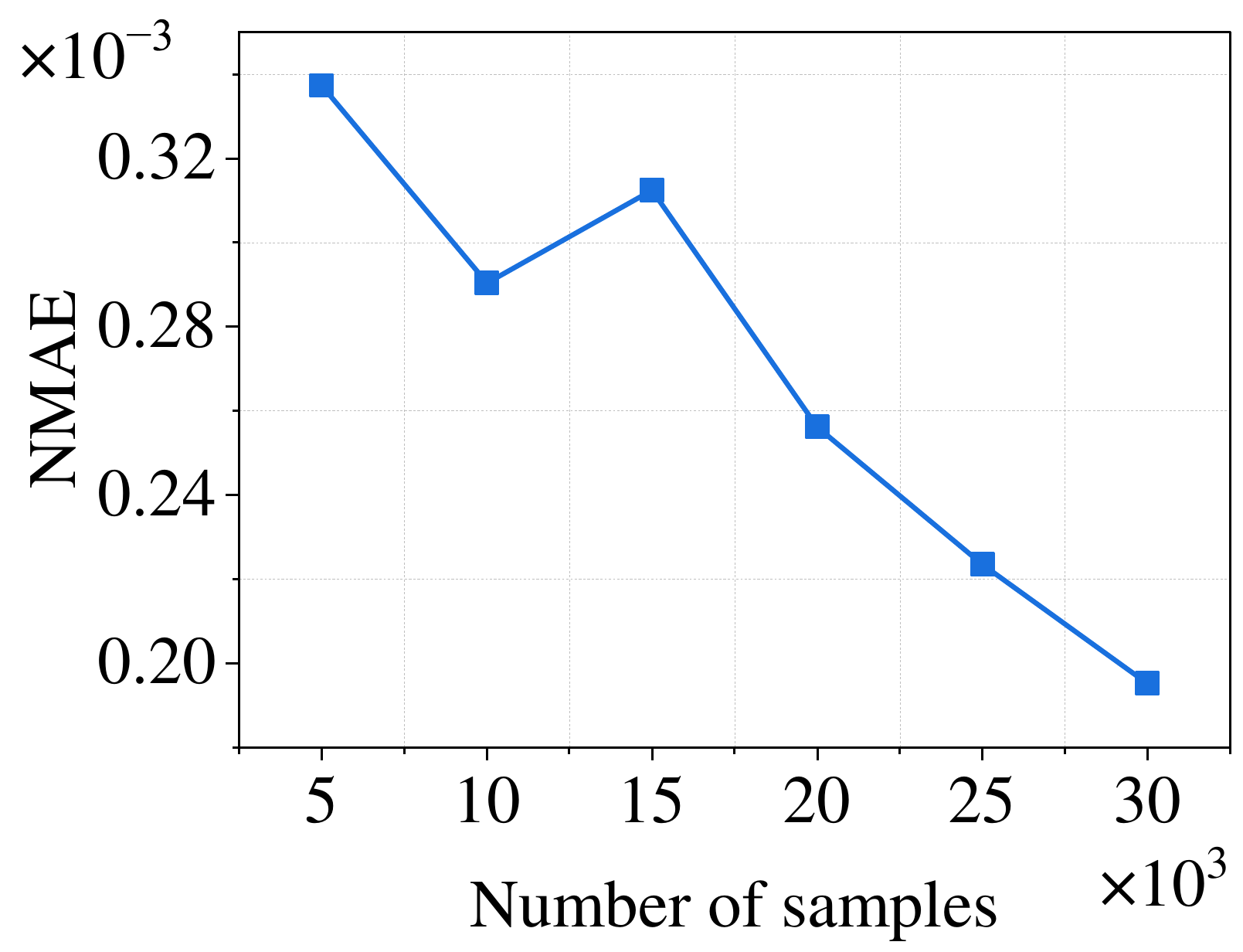}}\\
    \caption{\textcolor{black}{NMAE of ICNN outputs with different dataset sizes.}}
    \label{ICNN accuracy nmae}
\end{figure}

\textcolor{black}{
In practice, utilities typically have an exact model of their distribution feeder and are able to run power flow simulations under various operating conditions. Thus, it is possible for them to generate the required dataset using the same random sampling approach employed in this paper. Furthermore, if historical operational data is available, it can be leveraged to generate a training set that focuses more on common operating conditions. However, running power flow simulations to generate data is both time-consuming and memory-intensive. To shed light on the data volume requirement of ICNNs, we vary the dataset size from 5,000 to 30,000 power flow snapshots and compare the accuracy of ICNNs. As shown in Fig.~\ref{ICNN accuracy nmae}, as the dataset size increases, the NMAE decreases, indicating that ICNNs are becoming more accurate in replicating the distribution system power flow relationships. We also observe saturation effects as dataset sizes increase. As shown in Figs.~\ref{nmae loss} and \ref{nmae current}, the accuracy improvement becomes marginal when the dataset size exceeds 20,000. Considering all four ICNNs, we believe that a dataset of 30,000 samples strikes a balance between data generation efforts and ICNN accuracy requirements. Typically, more data samples are required when ICNNs are trained for larger distribution systems. Thus, the time required to generate samples could significantly increase. The sampling method proposed in Ref.~\cite{hu_optimal_2025} can be applied to large-scale systems to address this issue. Furthermore, it may be challenging to train a single ICNN to accurately represent the power flow of a large system. Proper partitioning of the system could be a solution to this issue.
}

\subsection{Effectiveness of ICNN in DOE Optimization}
The performances of the proposed method and the benchmarks are compared in terms of objective values and solution time. To showcase the advantages and disadvantages of each method, we split the objective (\ref{reformulated DOE obj}) into three terms:
\begin{align}
    \mathcal{J} = & \underbrace{w^{\text{DOE}}\sum_{j\in \mathcal{M}}{\left| p_{j}^{\max}-p_{j}^{+} \right|}}_{\mathcal{J}_1} +\underbrace{w^{\text{loss}}P^{\text{loss}}}_{\mathcal{J}_2} \notag\\
    &\quad\quad\quad\quad\quad\quad\quad\quad+ \underbrace{w^{\text{v}}\delta ^{\text{v}}+w^{\text{ol}}\delta ^{\text{ol}}+w^{\text{rpf}}\delta ^{\text{rpf}}}_{\mathcal{J}_3}.
\end{align}
Here, $\mathcal{J}_1$ denotes the deviation between the optimized DOEs and the original power limits of DERs; $\mathcal{J}_2$ quantifies the power loss of the system; $\mathcal{J}_3$ indicates constraint violations of the system. Since DOEs are used to ensure network integrity, we expect $\mathcal{J}_3$ to be minimized to zero, i.e., constraints (\ref{voltage limit}) - (\ref{reverse power flow limit}) are satisfied. In this context, we prefer DOEs that minimize the deviation from original DER power limits and reduce losses, i.e., the smaller $\mathcal{J}_1$ and $\mathcal{J}_2$, the better.

We first analyze the optimization results of one snapshot in detail. The results of optimizing $p_j^+$ and $p_j^-$ are summarized in Tables~\ref{p+ one snapshot} and \ref{p- one snapshot}, respectively. Benchmark B0 shows the result of unrestrained DER power. If DERs operate at $p_j^{\max}$, the reverse power flow constraint (\ref{reverse power flow limit}) will be violated, while if DERs operate at $p_j^{\min}$, voltage and overloading constraints will be violated, both resulting in large values of $\mathcal{J}_3$. Thus, it is necessary to calculate DOEs to constrain the power of DERs and avoid constraint violation. \textcolor{black}{Specifically, when applying benchmarks B1-B4, we optimize the problem to a solution where $J_3$ is zero, implying that there is no constraint violation. The last column of Tables~\ref{p+ one snapshot} and \ref{p- one snapshot} are the solution time of DOE optimization problems when different benchmark methods are applied.}

\begin{table}[bt]
\centering
\caption{Results of $p_j^+$ Optimization}
\label{p+ one snapshot}
\begin{tabular}{cccccc}
\toprule
   & $\mathcal{J}_1$ & $\mathcal{J}_2$ & $\mathcal{J}_3$ & $\mathcal{J}$ & Time (s) \\ \midrule
B0 & 0               & 14.16           & 927.52       & 941.68     & -        \\
B1 & 30.87          & 14.90           & 0               & 45.77        & 0.19     \\
B2 & 30.87          & 14.90           & 0               & 45.77        & 12.03    \\
B3 & 63.80          & 15.66           & 0               & 79.46        & 1.07     \\
B4 & 61.42          & 15.60           & 0               & 77.02       & 900 \\ \bottomrule
\end{tabular}
\end{table}

\begin{table}[bt]
\centering
\caption{Results of $p_j^-$ Optimization}
\label{p- one snapshot}
\begin{tabular}{cccccc}
\toprule
   & $\mathcal{J}_1$ & $\mathcal{J}_2$ & $\mathcal{J}_3$ & $\mathcal{J}$ & Time (s) \\ \midrule
B0 & 0               & 45.28           & 14633.43        & 14678.71     & -        \\
B1 & 170.75          & 35.14           & 0               & 205.90        & 0.20     \\
B2 & 170.75          & 35.14           & 0               & 205.90        & 12.10    \\
B3 & 257.06          & 30.80           & 0               & 287.86        & 1.05     \\
B4 & 428.20          & 27.44           & 0          & 455.64       & 900 \\ \bottomrule
\end{tabular}
\end{table}

\begin{figure}
    \centering
    \subfloat[Overall active power load of the test system.\label{overall load}]{%
       \includegraphics[width=\linewidth]{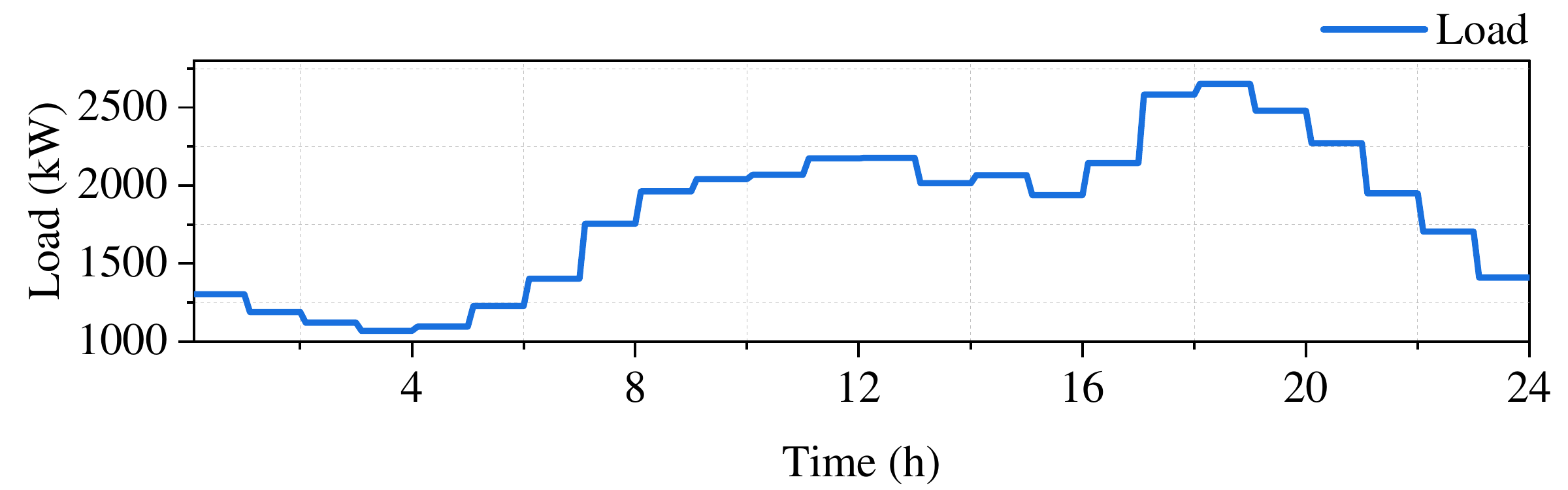}}\\
    \subfloat[DER 1.\label{p35}]{%
       \includegraphics[width=\linewidth]{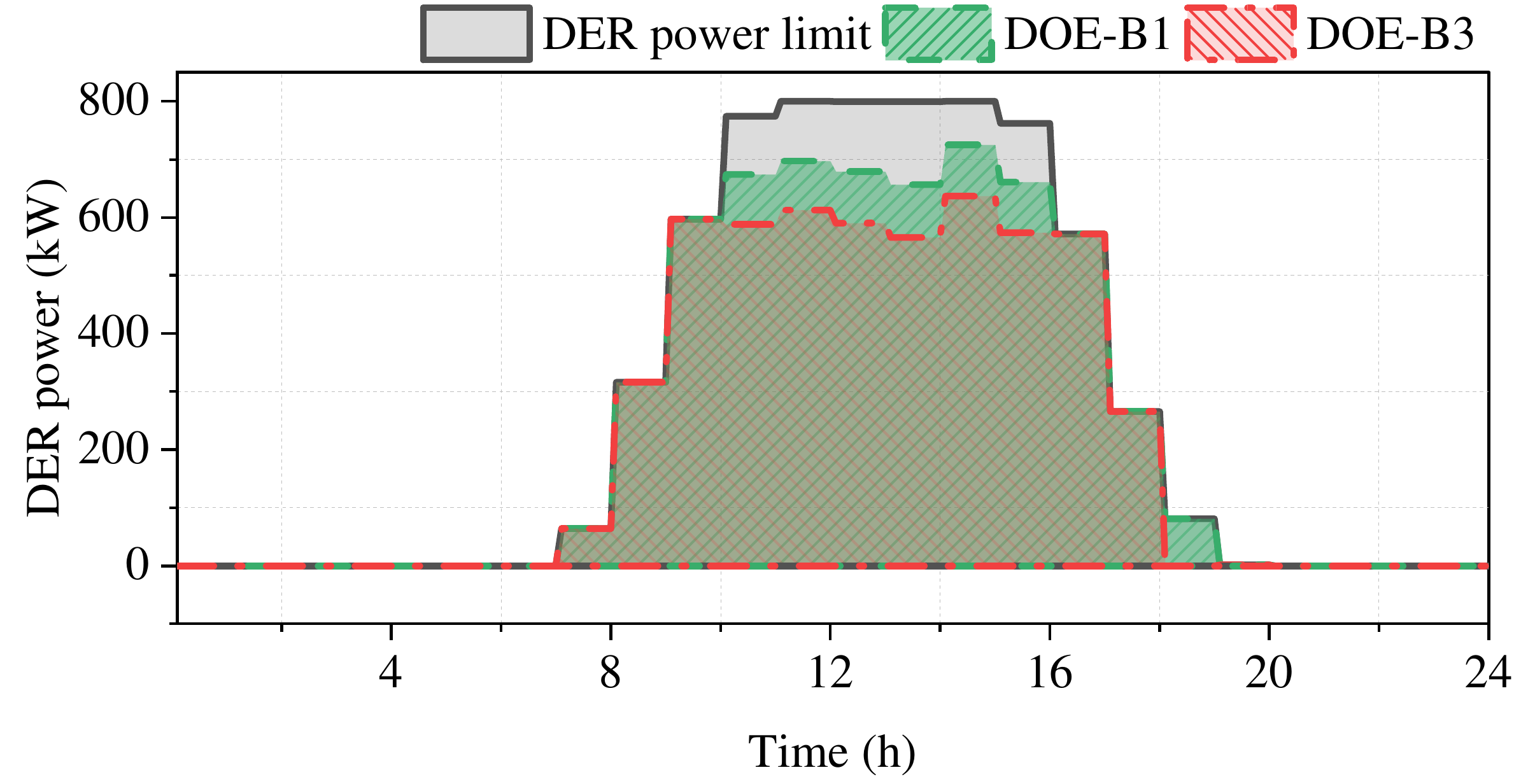}}\\
       \subfloat[DER 2.\label{p86}]{%
       \includegraphics[width=\linewidth]{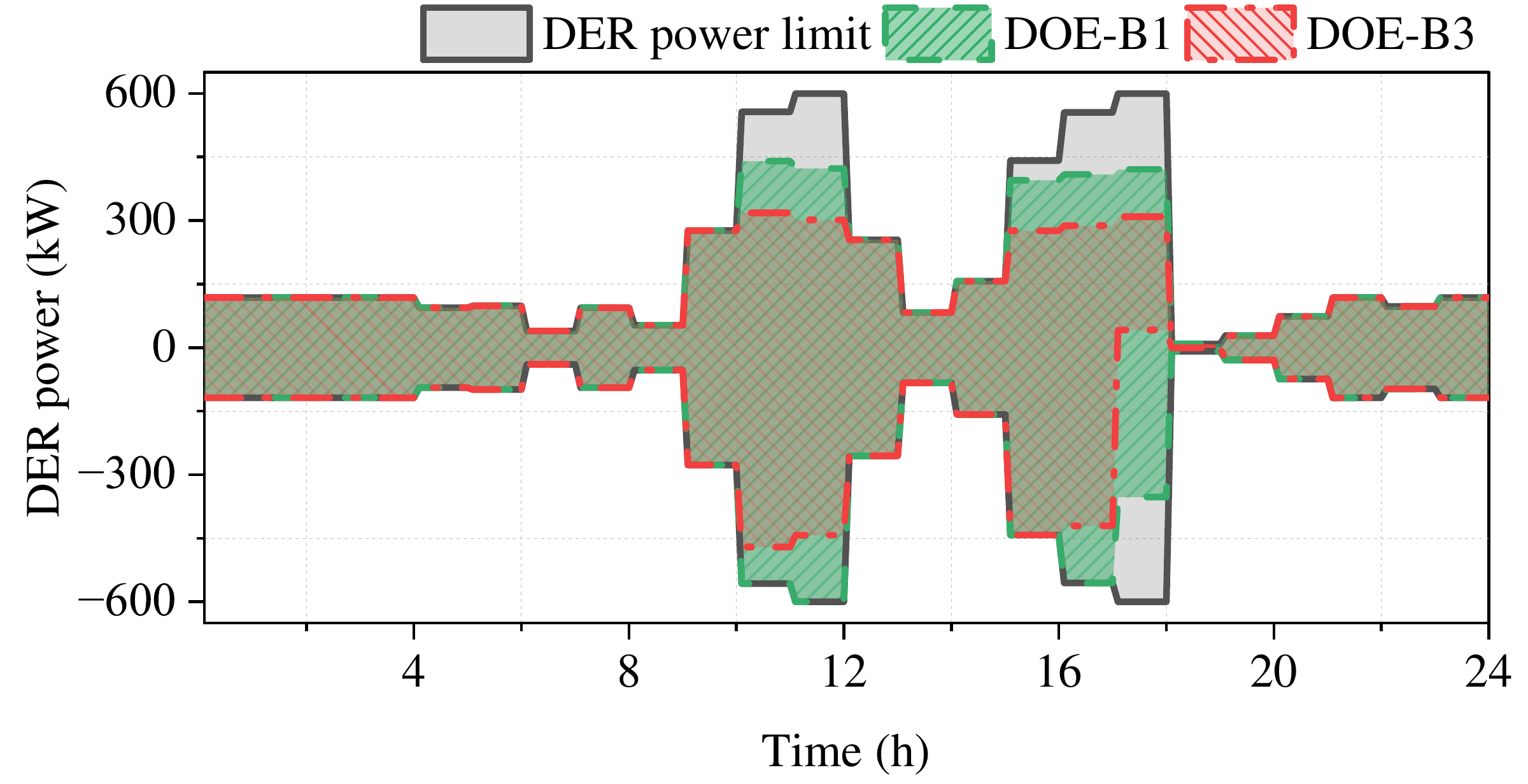}}
    \caption{Illustration of DOEs for DERs in one day.}
    \label{DOE illustration}
\end{figure}

B1 and B2 are ICNN-based methods proposed in this paper, which achieve the lowest objective value $\mathcal{J}$ in both $p_j^{+}$ and $p_j^{-}$ optimizations, outperforming the rest of the benchmark methods. Eliminating all constraint violations, B1 and B2 generate DOEs with the least deviation from $p_j^{\max}$ and $p_j^{\min}$ and minimal losses. The results demonstrate that the proposed ICNN-based methods can accurately capture the non-convex power flow relationships and effectively simplify the optimization of DOEs for DERs. Note that the solutions obtained by B1 and B2 are the same, implying that the proposed linear relaxation of ICNN is tight, as we theoretically proved in Theorem~\ref{ICNN DOE theorem}. With linear relaxation, the solution time of DOE optimization problem (\ref{reformulated DOE optimization}) reduces from more than 10 seconds (B2) to less than 0.3 seconds (B1), indicating that the proposed method significantly improves optimization efficiency.

As a model-based linearization method, the solution time of B3 is shorter than B2 but longer than B1. Although performing well in solution time, B3 is more conservative in issuing DOEs. The objective terms $\mathcal{J}_1$ and $\mathcal{J}_2$ of B3's results indicate larger deviations of DER power limits and less reduction in power loss. Thus, the DOEs generated by B3 are less optimal than the DOEs generated by B1 and B2. B4 is based on conventional neural networks, which can also accurately learn the power flow relationships, as demonstrated in Table~\ref{ICNN config table}. However, B4 needs to be reformulated into an MILP problem. In this paper, using the hidden layer configuration specified in Table~\ref{ICNN config table} results in an MILP with around 2,000 integer variables. In our experiments, Gurobi can only provide a sub-optimal solution of this non-convex MILP problem before it terminates at the 900-second solution time limit. In both cases, our proposed methods B1 and B2 significantly outperform B4 in terms of both solution optimality and solution time. 

We use hourly time-series data to further examine and compare the benchmarks. Since we have verified the equivalence of our proposed methods B1 and B2, we only show the results of B1 in the following discussions. B4 is excluded due to its excessively long and impractical solution time. The performance of B1 and B3 is compared in the context of no constraint violation, i.e., $\mathcal{J}_3=0$. The overall active power loads of the system and optimal DOEs of individual DERs in one day are visualized in Fig.~\ref{DOE illustration}. At night, since the load level and the DER power are both relatively lower, the DOEs issued by both B1 and B3 coincide with the original DER power limits. In the daytime, DOEs restrict both DERs from injecting power to the system at their upper bound to avoid reverse power flow. As shown in Fig.~\ref{p35} and Fig.~\ref{p86}, B3 is more conservative than B1 and sets lower upper limits for both DER 1 and DER 2. Since DER 1 is a PV, the lower limit of its DOE remains zero throughout the day. The power demand of DER 2 needs to be restricted by the lower limit of its DOE during peak hours, typically for the purpose of avoiding overloading of lines. We observe that B3 is overly conservative at some time intervals. For example, during hours 10-12, B3 increases the lower power limit of DER 2, which is unnecessary, while B1 is capable of issuing a more reasonable DOE to DER 2 without shifting its lower power limit during these time intervals. Overall, our proposed method B1 generates a wider DOE than B3, allowing DER aggregators to dispatch DERs more flexibly.

\subsection{Scalability of Proposed Method}
\begin{figure}
    \centering
    \subfloat[Objective $\mathcal{J}_1$: deviation between optimized DOEs
and original power limits.\label{power deviation}]{%
       \includegraphics[width=\linewidth]{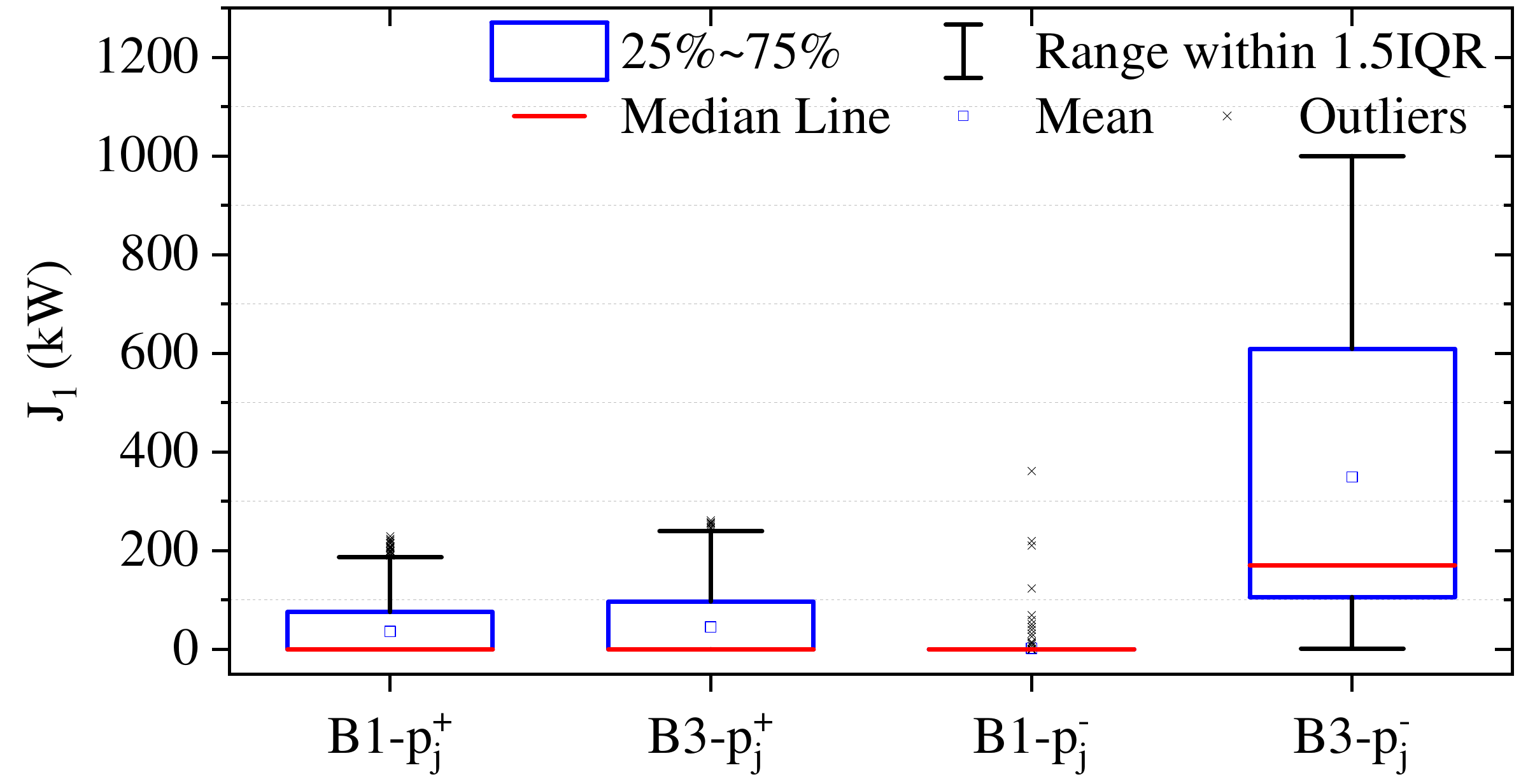}}\\
       \subfloat[Objective $\mathcal{J}_2$: power loss.\label{power loss}]{%
       \includegraphics[width=\linewidth]{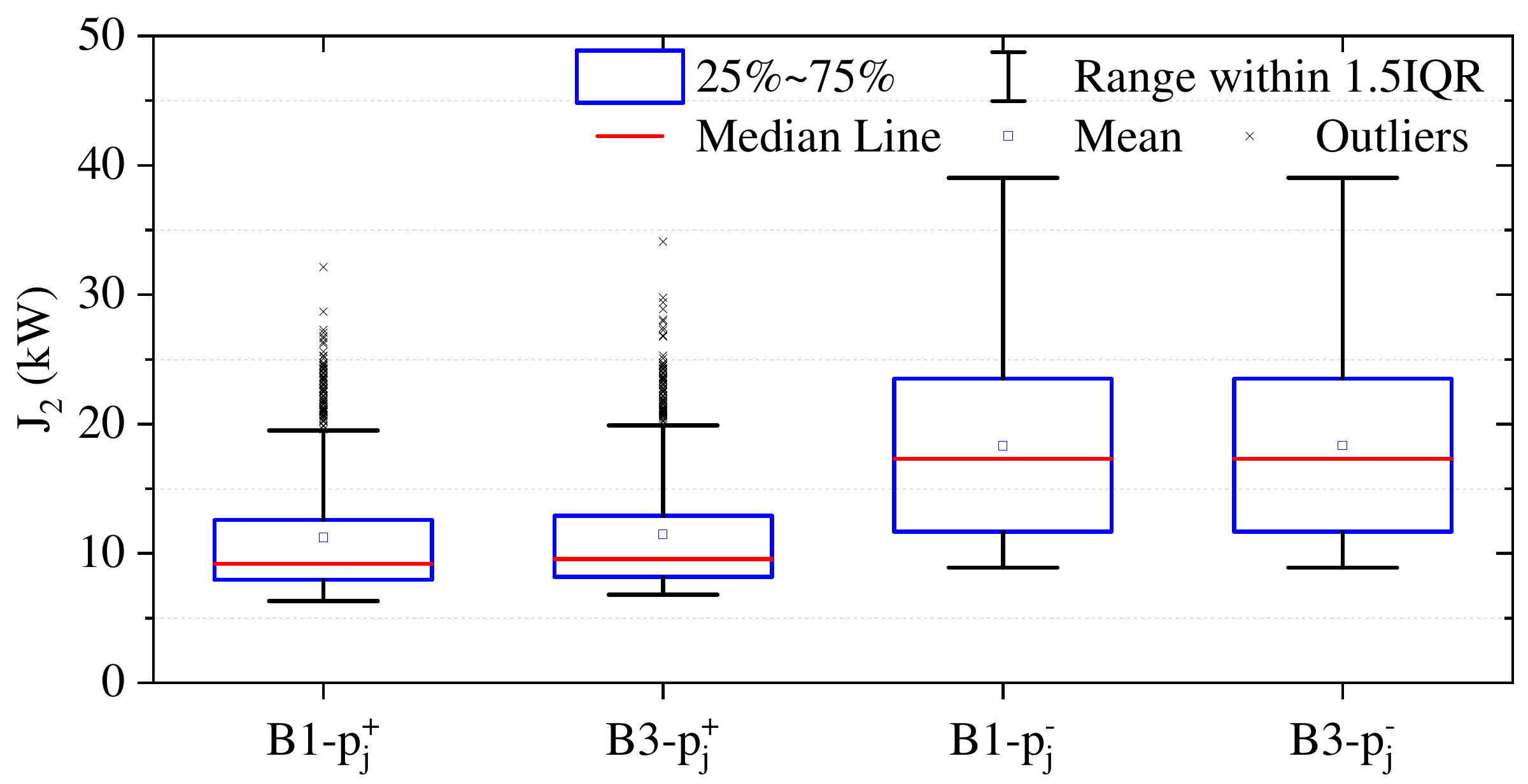}}\\
       \subfloat[Solution time.\label{solution time}]{%
       \includegraphics[width=\linewidth]{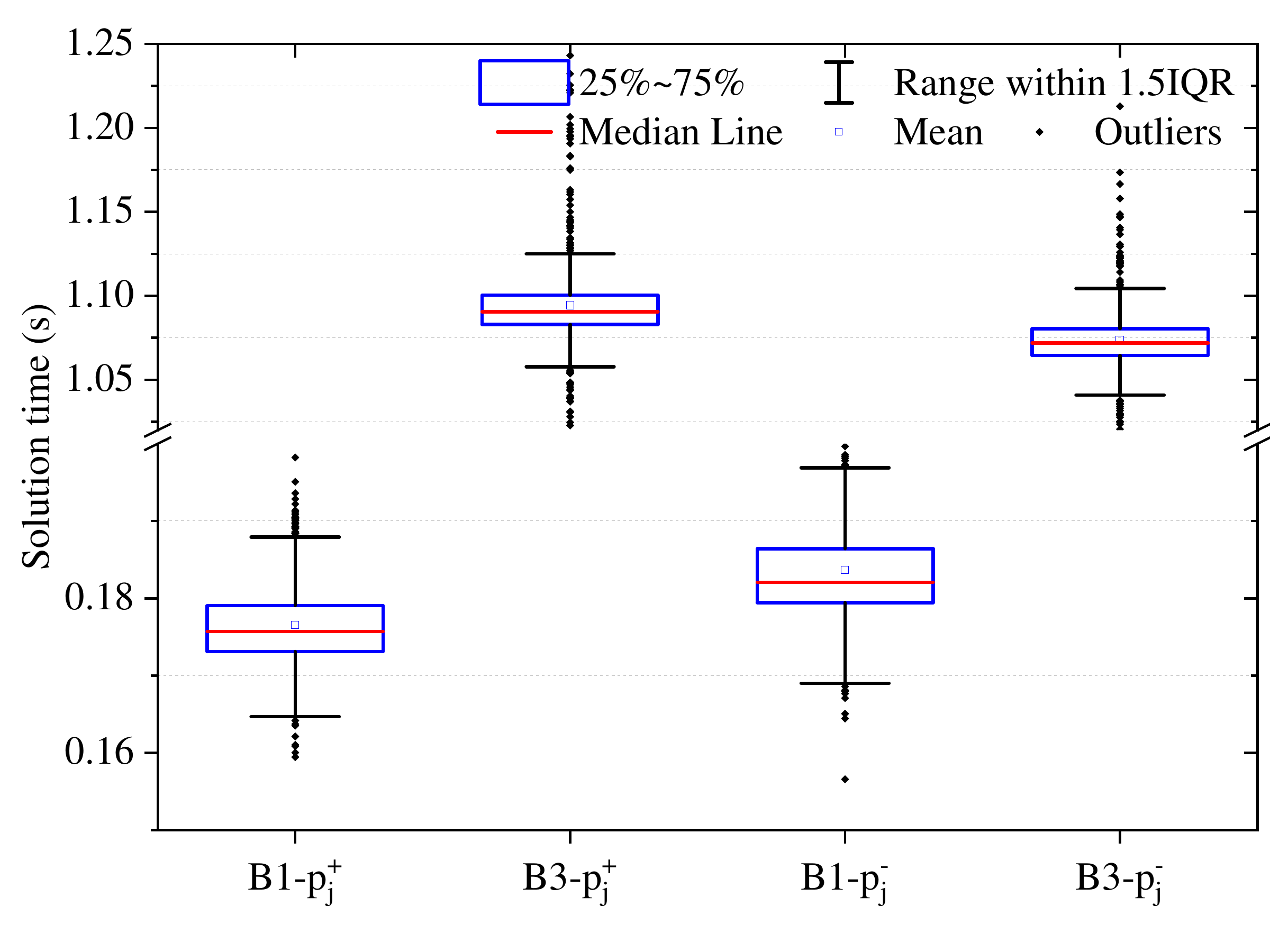}}
    \caption{Performance of benchmark methods over time-series data.}
    \label{comparison series data}
\end{figure}
\begin{figure}
    \centering
    \includegraphics[width=\linewidth]{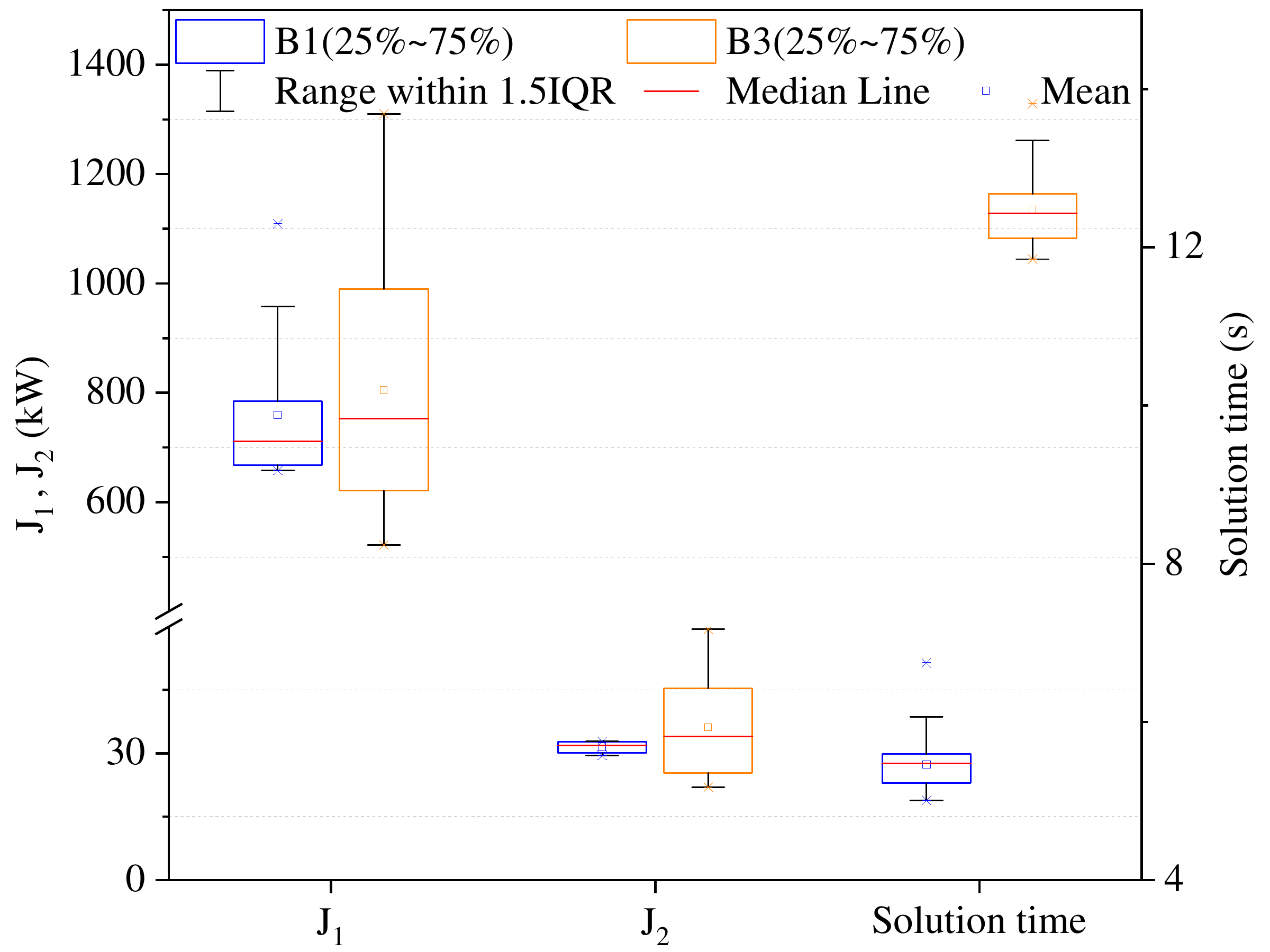}
    \caption{\textcolor{black}{Performance of benchmark methods on EPRI Ckt5 test system.}}
    \label{ckt5 performance}
\end{figure}
Fig.~\ref{comparison series data} compares the performance of B1 and B3 in optimizing hourly DOEs of 960 successive time intervals. Both benchmark methods generate DOEs without constraint violations. As shown in Fig.~\ref{power deviation}, our proposed method B1 is less conservative than B3 in issuing DOEs, for both $p^+_j$ and $p^-_j$. B1 significantly outperforms B3 in optimizing $p^-_j$, because B1 is more accurate in capturing power flow relationships. The performance of B1 and B3 in reducing power loss is similar, as shown in Fig.~\ref{power loss}. Thus, overall, B1 outperforms B3 in minimizing the objective of the DOE optimization problem (\ref{reformulated DOE optimization}), especially in $p^-_j$ cases. \textcolor{black}{As shown in Fig.~\ref{solution time}, the average solution time of B1 is 83.87\% and 82.90\% shorter than the solution time of B3.} In summary, compared to the LinDistFlow model B3, the proposed ICNN-based method B1 has higher accuracy in modeling power flow relationships and better computation efficiency in optimizing DOEs for DERs.

\textcolor{black}{
The performance of the benchmark methods B1 and B3 is further tested on the modified ERPI Ckt5 test system. Here, since the DERs are PV panels, only the upper bounds ($p_j^+$) of the DOEs are optimized, while the lower bounds ($p_j^-$) are kept to zero. This design avoids the generation obligation for PV panels due to weather uncertainty. The results of optimizing $p_j^+$ are shown in Fig.~\ref{ckt5}, showing a similar trend to those obtained from the modified IEEE 123-node test feeder. We compare the two components of the objective function, $J_1$ and $J_2$, respectively. $J_1$ is related to the adjustment of maximum PV outputs imposed by DOEs. As the results show, our proposed method B1 typically exhibits a lower $J_1$, indicating that with improved modeling accuracy, the proposed method forces less PV generation curtailment and promotes PV integration. The power loss $J_2$ of the optimal solutions of B1 is also lower on average, as more PV generation is utilized in the distribution system. The computational efficiency of B1 significantly outperforms B3, achieving a 52\% reduction in solution time. If we consider optimizing DOE upper and lower bounds for the next hour with a 5-minute granularity, it takes approximately 120-140 seconds to complete the optimization. Assuming that such DOE optimization is conducted every 5 minutes, there is still room for further security checks and result communication. Therefore, we conclude that the proposed method maintains an acceptable real-time performance as the system scales up. The high accuracy and computational efficiency of the proposed method indicate significant potential for facilitating DOE optimization in real-world applications.
}

\color{black}
\subsection{Sensitivity to Weighting Factors}
\begin{figure}
    \centering
    \includegraphics[width=\linewidth]{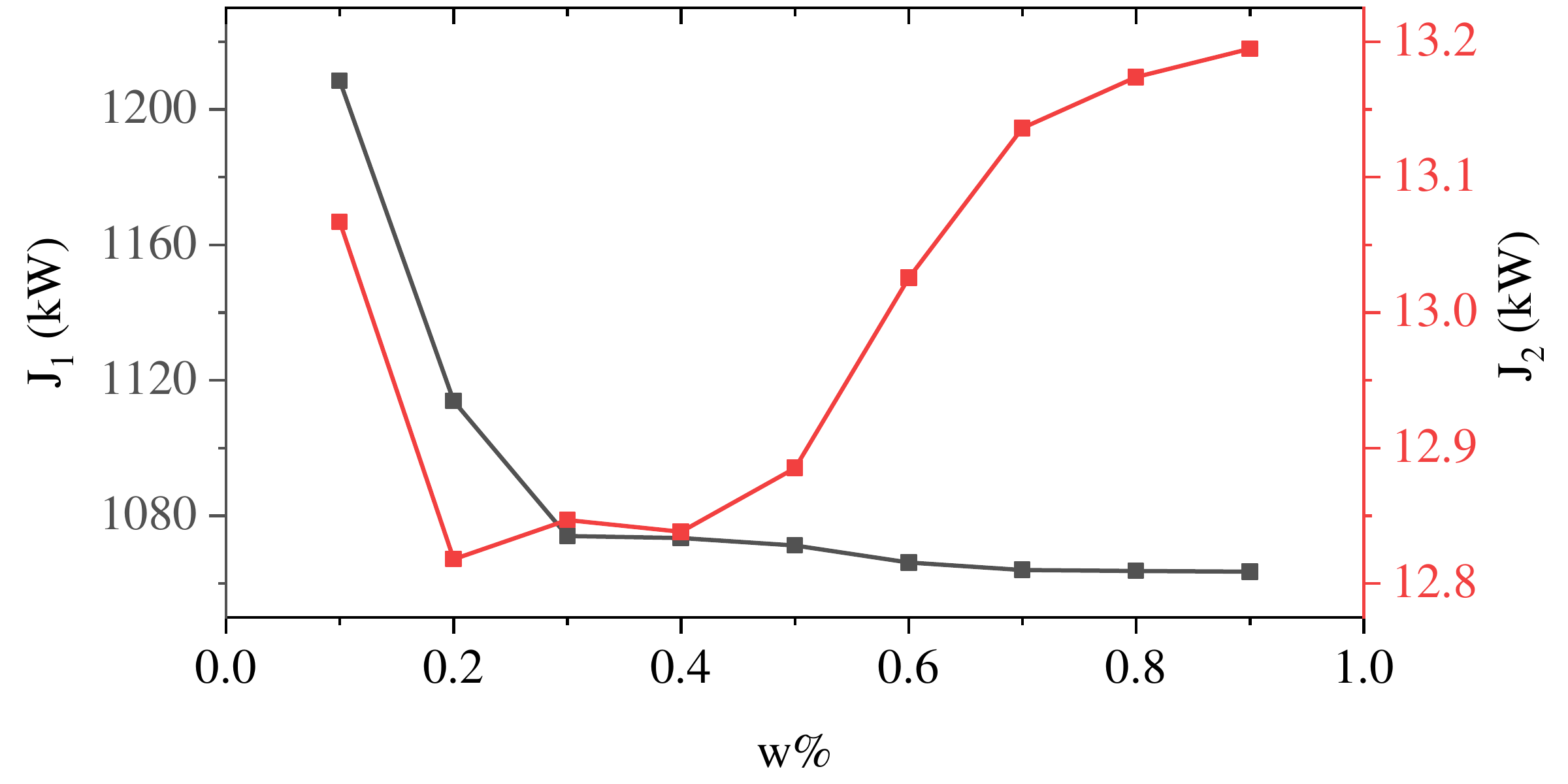}
    \caption{\textcolor{black}{Value of $J_1$ and $J_2$ under different weighting factors.}}
    \label{sensitivity}
\end{figure}
As discussed in the previous section, weighting factors $w_{\text{DOE}}$ and $w_{\text{loss}}$ can be adjusted to reflect the emphasize of DOE optimization. Here, we investigate the impact of different weighting factors on DOE optimization results. To maintain the equivalence between the penalized DOE optimization problem and the original problem, we keep the penalty factors $w^{\text{v}}$, $w^{\text{ol}}$, and $w^{\text{rpf}}$ large and unchanged to prevent constraint violations. The ratio between $w_{\text{DOE}}$ and $w_{\text{loss}}$ is defined as:
\begin{align}
    w\%=\frac{w_{\text{DOE}}}{(w_{\text{DOE}}+w_{\text{loss}})}.
\end{align}
We vary $w\%$ from 0.1 to 0.9 and solve the DOE optimization problem on the EPRI Ckt5 test system for each $w\%$. The results are shown in Fig.~\ref{sensitivity}. As $w\%$ increases, greater emphasis is placed on maximizing the size of the DOE, naturally resulting in a continuously decreasing value of $J_1$. The value of $J_2$ gradually increases as less emphasis is placed on reducing power loss. The only exception is the first two values, where $J_2$ decreases as $w\%$ increases from 0.1 to 0.2. The reason is that as $w\%$ increases from 0.1 to 0.2, the value of $J_1$ decreases significantly, implying that the size of the DOE is greatly enlarged. With a larger DOE, more PV generation is utilized to supply loads, which reduces power imports at the substation level and consequently lowers power loss across the system. 
\color{black}

\section{Conclusion}
\label{section 5}
In this paper, we exploit the convexity and linearity of ICNNs to optimize DOEs, thereby facilitating DER integration in distribution systems. First, we present the DOE optimization framework and develop a DOE optimization model that comprehensively considers multiple operation constraints. To address the non-convexity brought by power flow constraints, we reformulate the DOE optimization problem and train ICNN models to replicate power flow relationships. With the proposed constraint embedding method, the trained ICNN models are embedded into the reformulated DOE optimization problem, replacing the power flow constraints with mixed-integer counterparts. We further propose a linear relaxation of the ICNN-based DOE optimization model and prove the tightness of the relaxation. Through several case studies, we validate the accuracy of the ICNN models in replicating power flow equations and compare the performance of the proposed methods with benchmark methods. The results show that the proposed ICNN-based method excels in both solution accuracy and optimization efficiency. The tests on a large-scale system demonstrate the scalability of the proposed method, indicating significant potential for real-world applications. \textcolor{black}{Future research directions include: 1) confidence-aware and topology-adaptive ICNN models that can generalize to out-of-distribution operating conditions and topological variations; 2) sample-efficient and noise-resilient ICNN training technologies; 3) fair allocation of DOEs considering uncertainty of DERs and their pursuit for maximum grid access.}

\appendices
\section{LinDistFlow Formulation}
\label{LinDistFlow formulation}
Assuming that losses are negligible and defining $v_{j,t}=\left| V_{j,t} \right|^2$, equations (\ref{active power flow}) - (\ref{current definition}) are simplified to:
\begin{subequations}
\label{LinDistFlow}
    \begin{align}
&P_{ij,\phi ,t}=\sum_{k\in \mathcal{N} _{j}^{-}}{P_{jk,\phi ,t}}+p_{j,\phi ,t},\\
	&Q_{ij,\phi ,t}=\sum_{k\in \mathcal{N} _{j}^{-}}{Q_{jk,\phi ,t}}+q_{j,\phi ,t},\\
	&\left| V_{j,\phi ,t} \right|^2=\left| V_{i,\phi ,t} \right|^2-2\sum_{\psi \in \Phi}{\left( r_{ij,\phi \psi}P_{ij,\psi ,t}-2x_{ij,\phi \psi}Q_{ij,\psi ,t} \right)}.
\end{align}
\end{subequations}
Albeit linear, model (\ref{LinDistFlow}) does not explicitly model line current magnitudes and power losses. To cope with the DOE optimization problem (\ref{reformulated DOE optimization}, we propose the approximation of loss as:
\begin{align}
    P_{t}^{\text{loss}}=\sum_{(i,j)\in \mathcal{E}}{r_{ij}P_{ij,t}^{2}}+x_{ij}Q_{ij,t}^{2}.
\end{align}
The overloading constraint (\ref{thermal limit} is converted to:
\begin{align}
    \boldsymbol{P}_t\le \boldsymbol{P}^{\max},
\end{align}
where $\boldsymbol{P}^{\max}$ is empirically determined from $\boldsymbol{I}^{\max}$.

\bibliographystyle{IEEEtran}
\bibliography{references}

\end{document}